\documentclass[11pt]{elsarticle}

\usepackage{lineno,hyperref}
\modulolinenumbers[5]
\journal{Journal of XXX}

\usepackage{amsmath,amssymb,amsthm,amsfonts,afterpage,float,bm,stmaryrd,paralist,wrapfig,bbm,soul}

\usepackage{cancel}
\usepackage{color}
\usepackage{fancyhdr}
\usepackage{graphics}
\usepackage{hyperref}
\usepackage{graphicx,epsf}
\usepackage{makeidx}
\usepackage{epsfig}
\usepackage{lscape}

\usepackage{empheq,xcolor}

\pssilent

\numberwithin{equation}{section}
\numberwithin{figure}{section}
\numberwithin{table}{section}

\def\XXint#1#2#3{{\setbox0=\hbox{$#1{#2#3}{\int}$}
\vcenter{\hbox{$#2#3$}}\kern-.51\wd0}}

\graphicspath{{Figure/}}

\begin{document}

\setlength{\pdfpageheight}{\paperheight}
\setlength{\pdfpagewidth}{\paperwidth}
\title{Phase Field Modeling of  Dictyostelium Discoideum Chemotaxis}
\author{Yunsong Zhang}
\address{Department of Physics \& Astronomy and Center for Theoretical Biological Physics, Rice University, Houston TX, 77251-1892,USA}
\author{Herbert Levine}
\address{Center for Theoretical Biological Physics  \& Departments of Physics and Bioengineering, Northeastern University, Boston, MA 02115, USA}
\author{Yanxiang Zhao}
\address{Department of Mathematics, George Washington University, Washington D.C., 20052}

\begin{abstract}
A phase field approach is proposed to model the chemotaxis of Dictyostelium discoideum. In this framework, motion is controlled by active forces as determined by the Meinhardt model of chemical dynamics which is used to simulate directional sensing during chemotaxis. Then, the movement of the cell is achieved by the phase field dynamics, while the reaction-diffusion equations of the Meinhardt model are solved on an evolving cell boundary. This task requires the extension of the usual phase-field formulation to allow for components that are restricted to the membrane.  The coupled system is numerically solved by an efficient spectral method under periodic boundary conditions. Numerical experiments show that our model system can successfully mimic the typically observed pseudopodia patterns during chemotaxis.
\end{abstract}

\begin{keyword}
Phase field model; Chemotaxis; Dictyostelium discoideum;
\end{keyword}

\date{\today}
\maketitle

\section{Introduction}\label{Intro}

Many cells have an internal “compass”, which enables them to navigate through various environments. This “compass” detects the gradients in chemical concentrations, the rigidity of extra-cellular matrix, cellular adhesion sites, fluidic shear stress, etc. Interestingly, extensive studies on how such a “compass” is realized have led to “taxis-mania, focusing on chemotaxis, durotaxis, mechanotaxis, haptotaxis, plithotaxis and so on.

In this work, we specifically concentrate on chemotaxis, which plays an extensive role in many physiological processes \cite{BagordaParent_JCS2008}. For example, primordial cells are capable of figuring out their way to proper locations by sensing chemical clues, thus correctly forming the organs. Chemotaxis is also essential for immune responses and wound healing. In addition to these normal physiological processes, the pathology of numerous diseases, such as cancer metastasis and inflammatory disorders, is believed to be related to chemotaxis \cite{KedrinRheenen_JMGBN2007, Luster_NEJM1998}.

Dictyostelium discoideum, a type of amoeboid cell, is a popular model system for the study of chemotaxis. These cells rely on chemotaxis to find nutrition. When suffering from starvation, they are capable of chemotaxing in response to cAMP gradients in order to aggregate and enhance their chances for survival. Dicty chemotactic behavior is very similar to that of human leukocytes. Conceptually, chemotaxis can be divided into motility, directional sensing, and polarity \cite{DevreotesJanetopoulos_JBC2003,SwaneyHuangDevreotes,AndrewInsall_NatureCellBio007}. According to experiments \cite{AndrewInsall_NatureCellBio007,BosgraafHaastert_PLoSONE2009,XiongKabacoff_BMCSysBio2010}, motility involves periodic extensions and retractions of pseudopods -- temporary actin-filled protrusions of the cell membrane. Directional sensing refers to the process by which cells sense the chemical gradients and adjust their direction. Polarity refers tp the reorganization of the cell interior to favor moving in a  fixed direction. Once polarized, protrusions mainly extend from the cell anterior, regardless of whether a chemical gradient exists or not.

\subsection{Directional sensing: LEGI-BEN and Meinhardt models}

Detailed experimental investigations have uncovered many important features of chemotaxis \cite{IglesiasDevreotes_COCellBio2012}. First of all, the actin cytoskeleton in motile cells exhibits many characteristics of an excitable medium, such as the presence of propagating waves in the cell membrane \cite{VickerXiangPlathWosniok_PhysicaD1997,Vicker_BiophysChem2000,Vicker_ExpCellRes2002,Vicker_FEBSLetters2001}.This fact has suggested that an excitable network, which composes a simple activator-inhibitor system, may help explain the spontaneous migration of these cells. Then, to include the cell’s response to external gradients, one can modify the activator-inhibitor system by adding a steering bias: higher concentrations of chemoattractants will lower the threshold of excitability, thus causing more excitation. Over time, cells with this bias will tend to move along the directions towards higher chemoattractant density. Chemotactic cells can display surprising sensitivity, responding to a chemical gradient as small as 1\%. Such sensitivity can be captured by a biased excitable network \cite{Hecht_PLosComptBio2011}.

Although successful in qualitatively explaining spontaneous cell motion in the absence of chemoattractants as well as directed motion in the presence of chemoattractants, the biased excitable network (BEN) approach still misses some features of realistic chemotactic behavior. One such feature is the adaptability of cells to external chemical clues. Since higher concentrations of chemoattractants lower the threshold for excitable behavior according to a biased excitable network,  the cells would be predicted to become "hyper-excited" if they are exposed to a spatially uniform increase in chemoattractant concentration  In fact, this does not occur and after transient responses to uniform increments in chemoattractants, the excitability returns to baseline levels.  To account for this behavior, a local-excitation, global-inhibition (LEGI) mechanism was proposed \cite{ParentDevreotes_Science1999,LevchenkoIglesias_Biophys2002}. According to this mechanism, chemoattractants give rise to the release of a slowly diffusing activator which is accompanied by a rapidly diffusing inhibitor. Thanks to the regulation by the global inhibitor, the level of excitation returns to a threshold level.

Thus, one natural strategy is to directly combine the LEGI mechanism and the biased excitable network into a hybrid LEGI-BEN model \cite{XiongHuangIglesiasDevreotes_PNAS2010}. One can however use a different model which already contains the two essential components of excitation and adaptation. More than a decade ago, Meinhardt proposed a three-component model \cite{Meinhardt_JCS1999}, which shares a similar conceptualization with the LEGI-BEN approach. Here, a biased excitable network, which includes a slowly diffusing activator and a fast inhibitor, is further regulated by an extra even faster global inhibitor. The total quantity of the activator is held approximately constant over time by this global inhibitor, thus preventing a large increase in excitable behavior. For the sake of mathematical simplicity, we choose to use the Meinhardt model in our modeling efforts, whereas the LEGI-BEN model can also in the future be embedded in our phase field model, if needed.

The governing equations of Meinhardt model are:
\begin{align}
\frac{\partial a}{\partial t} &= D_a \nabla^2 a + r_a\frac{s(\textbf{r},t)(a^2b^{-1}+b_a)}{(s_c+c)(1+s_aa^2)} - r_a a, \\
\frac{\partial b}{\partial t} &= \frac{r_b}{|\Gamma(b)|} \oint a \text{d}x - r_b b, \\
\frac{\partial c}{\partial t} &= D_c \nabla^2c + b_c a - r_c c.
\end{align}
Here, $a, b$ and $c$ are respectively the local activator, the global inhibitor, and the local inhibitor. By referring to $c$ as a global inhibitor, we mean $D_b \gg D_c$, so that the inhibiting effects of $b$ can spread over the whole interface in almost no time, thus regulating the total quantity of the activator. Therefore we can assume the concentration of $b$ to be uniform everywhere on the membrane, which leads to the replacement of partial differential equation by an ordinary differential equation with a nonlocal source. This system exhibits nice bifurcating patterns \cite{Meinhardt_JCS1999}, which have been successfully compared in experimental findings \cite{Neilson_PLosBio2011}. A local excitation bifurcates into a pair of competing daughter bursts of excitations, which travel in opposite directions. One of these daughter bursts will win out over the other, which vanishes. A new bifurcation will occur once the “loser” dies. The decisive factor for the competition between the pair of bifurcations is the spatial factor $s(\textbf{r}, t)$ in the excitation of the activator:
\begin{align}
s(\textbf{r},t) = (1+d_r\xi)(1+C_{\text{chem}}f(\textbf{r},t)),
\end{align}
where $f(\textbf{r},t)$ is a function of the spatial clue related to the concentrations of chemoattractant. The factor $(1+d_r\xi)$ represents the effect of stochastic fluctuations, with $\xi$ taken as white noise and $d_r$ being the fluctuation strength.  In this work, we assume the existence of a chemoattractant source $\textbf{r}_0$, and simply set $f(\textbf{r},t)$ as:
\begin{align}
f(\textbf{r},t) = 1 - \frac{\text{dist}(\textbf{r},t) - \text{dist}_{\min}(t)}{\text{dist}_{\max}(t) - \text{dist}_{\min}(t)}
\end{align}
where $\text{dist}(\textbf{r},t)$ represents the distance between any point $\textbf{r}$ and the chemoattractant source $\textbf{r}_0$, while $\text{dis}_{\max}(t)$ and $\text{dis}_{\min}(t)$ respectively represent the maximal and minimal distance at time $t$, between the chemoattractant source $\textbf{r}_0$ and all points on the cell membrane. It is evident that $f(\textbf{r},t)$ varies between $0$ and $1$, monotonically decreasing with the distance from the chemoattractant source $\textbf{r}_0$, along the cell membrane. The parameter $C_{\text{chem}}$ regulates the strength of the bias on the threshold of the activator’s excitability. It turns out that the values of $C_{\text{chem}}$ as small as 0.01, can still significantly affect the bifurcation patterns in space and time. This fact is consistent with the chemotactic cells’ sensitivity to chemical gradients as mentioned above. Statistically,  excitation bursts propagating toward favorable positions in the chemical gradients are more likely to survive and continue to bifurcate, thus fostering the directed navigation of chemotactic cells.

\subsection{Phase field model framework}

In the past few decades, phase field models have emerged as one of the most successful methods for studying interfacial problems; see the two review articles \cite{Chen_ARMR2002, DuFeng_BookChapter2020} and the references therein. In the phase field model framework, a phase field function $\phi$ is introduced, assigning a value (say, 0) for one phase, and another value (say, 1) for the other. In the interfacial region, the phase field $\phi$ rapidly but smoothly transitions from 0 to 1. The interface is tracked by the 1/2-level set during the morphological evolution. The main advantage of the phase field approach is that it can allow for the computation of the temporal evolution of arbitrary morphologies and complex microstructures without explicitly tracking interfaces.

In the realm of biology, cell shape dynamics and cell migration processes have been simulated by using phase field models. In \cite{Shao_PRL2010}, a quantitative model for cell shape and motility dynamics  was constructed based on the original phase field concept \cite{CollineLevine_PRB1985}. An auxiliary field is introduced to distinguish the cell's interior ($\phi = 0$) from the exterior ($\phi = 1$). The dynamics of the cell are governed by equations that couple this field to the actual physical degrees of freedom, and the diffuse layer separating the interior from the exterior marks the membrane location. 

This cell motility model is part of a larger set of recent theoretical studies that have attempted to model cell migration. For example, some studies have attempted to calculate the "flow" of the actin cytoskeleton in a one-dimensional \cite{Gracheva_BullMathBio2004, Larripa_PhysicaA2006, Carlsson_NewJPhys2011} or fixed two-dimensional cell geometry \cite{Rubinsten_BiophysJ2009}. In some works, the cell boundary was allowed to change according to a phenomenological function of protrusion rate \cite{Barnhart_PLosBiol2011, Wolgemuth_BiophysJ2011} while other approaches implemented physical forces along the cell membrane, obtained cell shape and speed, but ignored actin flow and detailed adhesion mechanisms \cite{Shao_PRL2010, Ziebert_JRSocInterface2011}. Yet others examined adhesion dynamics and cell-substrate coupling while ignoring cell deformations \cite{Buenemann_BiophysJ2010} or focused on the dynamics of the leading edge \cite{Zimmermann_PRE2010}. Ziebert, Aranson and their coworkers studied the cell shape dynamics by coupling a vector field model of the actin filament network with the cell shape \cite{Ziebert_PLosOne2013}. Finally, a more comprehensive model for cell migration was presented in \cite{Shao_PNAS2012} which couples actin flow with discrete adhesion sites and deformable cell boundaries. Other interesting patterns such as periodic migration \cite{Camley_PRL2013} or circular motion \cite{Camley_PNAS2014, Camley_PRE2017} have also been studied using phase field framework.

\section{Phase field model of Chemotaxis}

\subsection{Pseudopodia: Chemical dynamics on a phase field membrane}
Our goal in this paper is to couple the aforementioned directional sensing system to a computational model of the resultant motion.
With the intricate spatio-temporal patterns of the Meinhardt model in hand, our immediate challenge is how to make this happen on the membrane of a dynamically evolving cell described by a phase field. To our best knowledge, there are very few published attempts to study similar problems. For example, Nelson et al. applied a hybrid computational framework to couple the Meinhardt model with cell movement \cite{Neilson_PLosBio2011}. There, the movement of the cell is achieved using a level set method \cite{Osher_Book2003}, while the reaction-diffusion equations of the Meinhardt model are approximated on an evolving cell boundary using an arbitrary Lagrangian-Eulerian surface finite element method (ALE-SFEM). In our phase field model, we can achieve the same effect in a much simpler manner. To accomplish this, we modified the approach used to couple bulk reaction-diffusion systems to phase field cells \cite{Shao_PRL2010, Camley_PNAS2014}. Specifically, instead of a factor of $\phi$ to limit reaction to the interior, we use $g(\phi(\textbf{r})) = \frac{\epsilon}{2}|\nabla\phi|^2$ to restrict concentrations to the membrane (see below). In addition, we found it necessary to insert diffusion in the normal direction of the phase field interface, so that reaction-diffusion processes in different layers synchronize with each other. Our revised equations for the membrane Meinhardt system are:
\begin{align}
&\tau_0 \frac{\partial(g(\phi)a)}{\partial t} + \nabla\cdot(g(\phi)a\mathbf{v}) = D_a\nabla_{\parallel}\cdot(g(\phi)\nabla_{\parallel}a) + D_{\perp}\nabla_{\perp}\cdot(g(\phi)\nabla_{\perp}a) \nonumber\\
&\hspace{2.5in}+ g(\phi)\left( \frac{s(\textbf{r},t)(a^2b^{-1}+b_a)}{(s_c+c)(1+s_aa^2)} - r_a a \right), \label{eqn:a}\\
&\tau_0 \frac{\partial b}{\partial t}  =  r_b\frac{\int g(\phi)a\text{d}\textbf{r}}{\int g(\phi)\text{d}\textbf{r}} - r_b b , \label{eqn:b}\\
&\tau_0 \frac{\partial(g(\phi)c)}{\partial t} + \nabla\cdot(g(\phi)c\mathbf{v}) = D_c\nabla_{\parallel}\cdot(g(\phi)\nabla_{\parallel}c) + D_{\perp}\nabla_{\perp}\cdot(g(\phi)\nabla_{\perp}c) + g(\phi)\left( b_c a - r_c c \right). \label{eqn:c}
\end{align}
where $\mathbf{v}$ is the interface velocity equal to $- \partial_t \phi \frac{\mathbf{\nabla} \phi}{| \mathbf{\nabla} \phi |^2}$. As already mentioned, the global inhibitor $b$ immediately spreads over the whole membrane, and it satisfies an ordinary differential equation instead of a partial differential equation. The term $g(\phi(\textbf{r})) = \frac{\epsilon}{2}|\nabla\phi|^2$ is only nonzero in the interfacial region so that the reaction-diffusion dynamics only occur near the interface, and $D_{\perp}$ refers to the diffusion we add in the normal direction of the membrane. More specifically, $\nabla_{\parallel}$ and $\nabla_{\perp}$ read:
\begin{align}\label{eqn:par_perp}
\nabla_{\parallel} = 
\begin{bmatrix*}[c]
      n_y^2   & -n_xn_y  \\
      -n_xn_y    & n_x^2 \\
\end{bmatrix*}
\begin{bmatrix*}[c]
      \partial_x  \\
      \partial_y \\
\end{bmatrix*},
\quad 
\nabla_{\perp} = 
\begin{bmatrix*}[c]
      n_x^2   & n_xn_y  \\
      n_xn_y    & n_y^2 \\
\end{bmatrix*}
\begin{bmatrix*}[c]
      \partial_x  \\
      \partial_y \\
\end{bmatrix*},
\end{align}
in which the normal vector $\textbf{n} = [n_x, n_y]^T$ can be calculated by $\textbf{n} = -\frac{\nabla\phi}{|\nabla\phi|}$. In practice, $\nabla_{\perp}$ has to be much larger than the other diffusion coefficients in the chemical systems. The detailed parameters used in our model simulations are listed in Table \ref{table:parameter}.

\begin{table}
\begin{center}
\begin{tabular}{ |c|c|c|c| } 
 \hline
Parameter & Value & Parameter &Value \\ 
$D_a$ & 8e-3 & $r_a$ & 0.2 \\ 
$D_c$ & 1.8$D_a$ & $r_b$ & 0.3 \\ 
$\nabla_{\perp}$ & 1.0 & $r_c$ & 0.13 \\ 
$s_a$ & 5e-4 & $b_a$ & 0.1 \\ 
$\tau_0$ & 0.01 & $b_c$ & 0.05 \\ 
$\tau_0$ & 0.01 & $s_c$ & 0.2 \\ 
$dr$ & 0.02 & $C_{\text{chem}}$ & 0.02 \\ 
 \hline
\end{tabular}\label{table:parameter}
\caption{Parameters in the Meinhardt model coupled with a phase field membrane.}
\end{center}
\end{table}

\subsection{Chemotaxis dynamics of Dictyostelium discoideum}

We model the Dictyostelium discoideum cell as a 2d region with a fixed area $A_0$. The evolving shape of the cell membrane is determined by the competition of several forces, including surface tension, bending force, the pressure that constrains the cell area, the chemical protrusive force which is proportional to the density of local activator $a$, and the effective friction due to the interaction between cell membrane and the substrate. All of them are formulated under the phase field framework, as follows. 

Given the surface energy in phase field formulation \cite{Du_JCP2006}:
\begin{align}\label{energy:tension}
E_{\text{ten}} = \gamma \int_{\Omega} \left( \frac{\epsilon}{2}|\nabla \phi|^2 + \frac{1}{\epsilon}G(\phi)  \right) \text{d}\mathbf{r},
\end{align}
in which $\gamma$ is the surface tension, $\epsilon$ is the phase field parameter controlling the width of the cell membrane (the width of phase field interface), and $G(\phi) = 18\phi^2(1-\phi)^2$ is a double well potential with minima at $\phi=0$ and $\phi = 1$, the surface tension is derived by taking the variational derivative of the surface energy \cite{Shao_PRL2010},
\begin{align}
\mathbf{F}_{\text{ten}} = \frac{\delta E_{\text{ten}}}{\delta\phi}\frac{\nabla\phi}{\epsilon|\nabla\phi|^2} = \frac{\gamma}{\epsilon}\left( -\epsilon \nabla^2\phi + \frac{1}{\epsilon}G'(\phi)  \right) \frac{\nabla\phi}{|\nabla\phi|^2}.
\end{align}
Similarly, given the bending energy in phase field formulation \cite{Du_JCP2006}:
\[
E_{\text{bend}} = \frac{\kappa}{2} \int_{\Omega} \frac{1}{\epsilon}\left(  \epsilon \nabla^2\phi - \frac{1}{\epsilon}G'(\phi)   \right)^2 \text{d}\mathbf{r},
\]
with $\kappa$ being the bending rigidity, we obtain the bending force \cite{Shao_PRL2010}:
\begin{align}
\mathbf{F}_{\text{bend}}  = \frac{\delta E_{\text{bend}}}{\delta\phi}\frac{\nabla\phi}{\epsilon|\nabla\phi|^2} = \frac{\kappa}{\epsilon^2}\left(  \epsilon\nabla^2 - \frac{1}{\epsilon}G''  \right) \left(  \epsilon\nabla^2\phi - \frac{1}{\epsilon}G'  \right) \frac{\nabla\phi}{|\nabla\phi|^2}.
\end{align}
The area force is given as a soft penalty on cell area:
\begin{align}
\mathbf{F}_{\text{area}} = M_{\text{area}} \left( \int_{\Omega} \phi \text{d}\mathbf{r} - A_0  \right) \frac{\nabla\phi}{|\nabla\phi|},
\end{align}
in which $M_{\text{area}}$ is the penalty constant. We assume the protrusion force is simply proportional to the density of the local activator $a$ on the membrane. Given the fact that the activator’s concentration may change over several magnitudes, we added a saturating restriction to the force $|\mathbf{F}_{\text{chem}}| \propto \tilde{\alpha} = \max(10,a)$,
\begin{align}
\mathbf{F}_{\text{chem}} = -\alpha \tilde{a} \frac{\nabla\phi}{|\nabla\phi|},
\end{align}
in which $\alpha$ is the strength of the chemical protrusion force. One could alternatively use a sigmoidal function to achieve the same effect. A friction force due to the interaction between the cell and the substrate (such as adhesion, attachment and detachment of the cell from the substrate) is introduced which is proportional to the local speed: $\mathbf{F}_{\text{fr}} = -\tau\mathbf{v}$. The force balance at quasi-steady state 
\[
\mathbf{F}_{\text{tot}} = \mathbf{F}_{\text{ten}} + \mathbf{F}_{\text{bend}} + \mathbf{F}_{\text{area}} + \mathbf{F}_{\text{chem}} + \mathbf{F}_{\text{fr}} = 0
\]
yields
\[
\mathbf{v} = -\frac{1}{\tau}\mathbf{F}_{\text{fr}} = \frac{1}{\tau}(\mathbf{F}_{\text{ten}} + \mathbf{F}_{\text{bend}} + \mathbf{F}_{\text{area}} + \mathbf{F}_{\text{chem}}).
\]
Finally using the transport equation of the phase field $\phi$ along the velocity field $\mathbf{v}$: $\frac{\partial\phi}{\partial t} + \mathbf{v}\cdot\nabla\phi = 0$, we obtain the following equation for $\phi$:
\begin{align}\label{eqn:phi}
\tau\frac{\partial\phi}{\partial t} = &-\kappa \left(\nabla^2 - \frac{G''(\phi)}{\epsilon^2}\right)\left( \nabla^2\phi - \frac{G'(\phi)}{\epsilon^2}  \right) + \gamma \left( \nabla^2\phi - \frac{G'(\phi)}{\epsilon^2}  \right) \nonumber\\
&- M_{\text{area}}\left( \int \frac{\epsilon}{2}|\nabla\phi|^2 + \frac{1}{\epsilon}G(\phi) \text{d}\textbf{r} - P_0 \right)|\nabla\phi| + \alpha \tilde{a}|\nabla\phi|.
\end{align}
Physically, this chemotaxis dynamics of $\phi$ implies that the friction force on the cell is balanced by the chemical protrusion force, which is transmitted from the substrate onto the cell via adhesion complexes.

\section{Numerical Simulations}

\begin{figure}[t] 
\begin{center}
\includegraphics[width=0.24\linewidth]{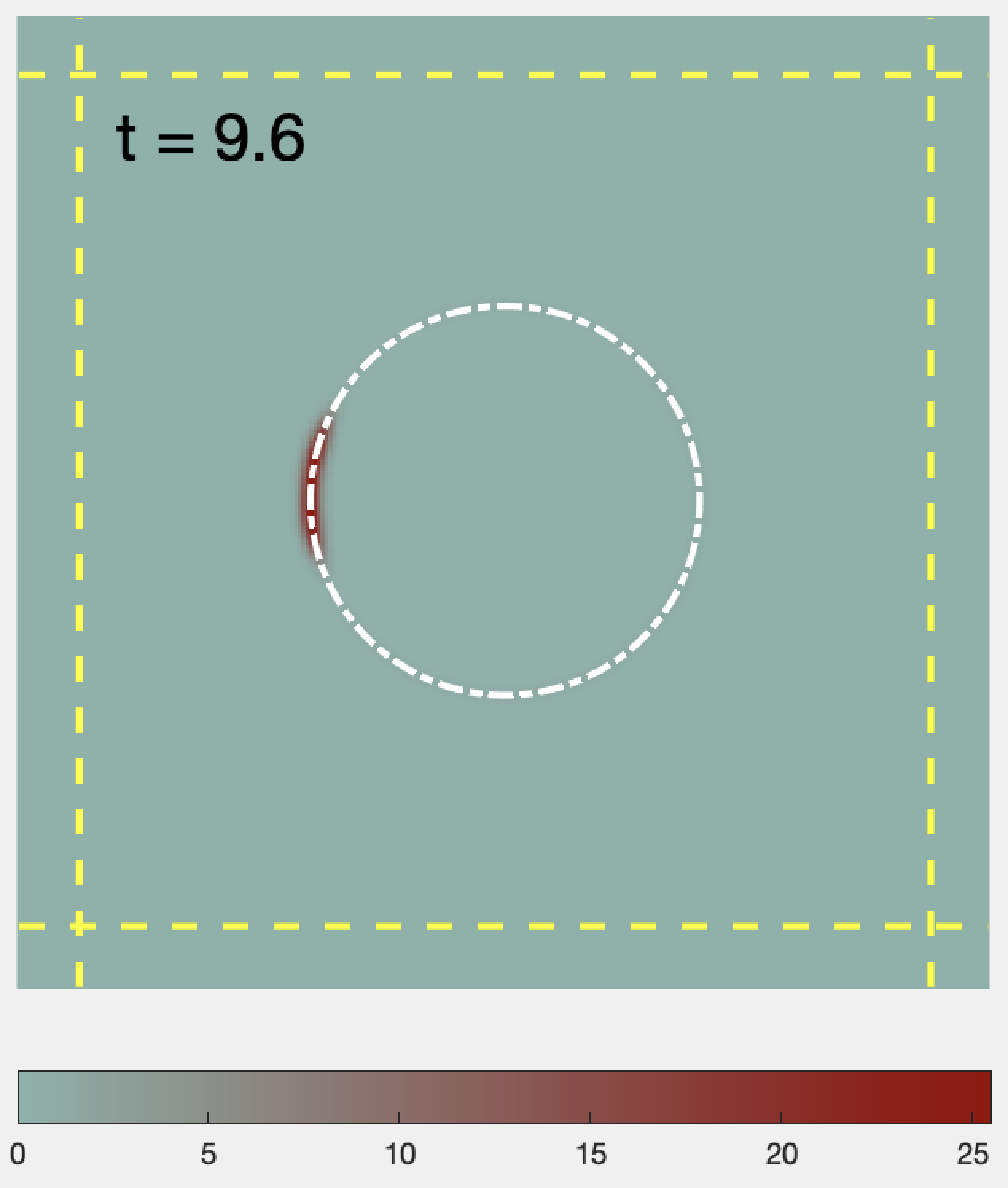}
\includegraphics[width=0.24\linewidth]{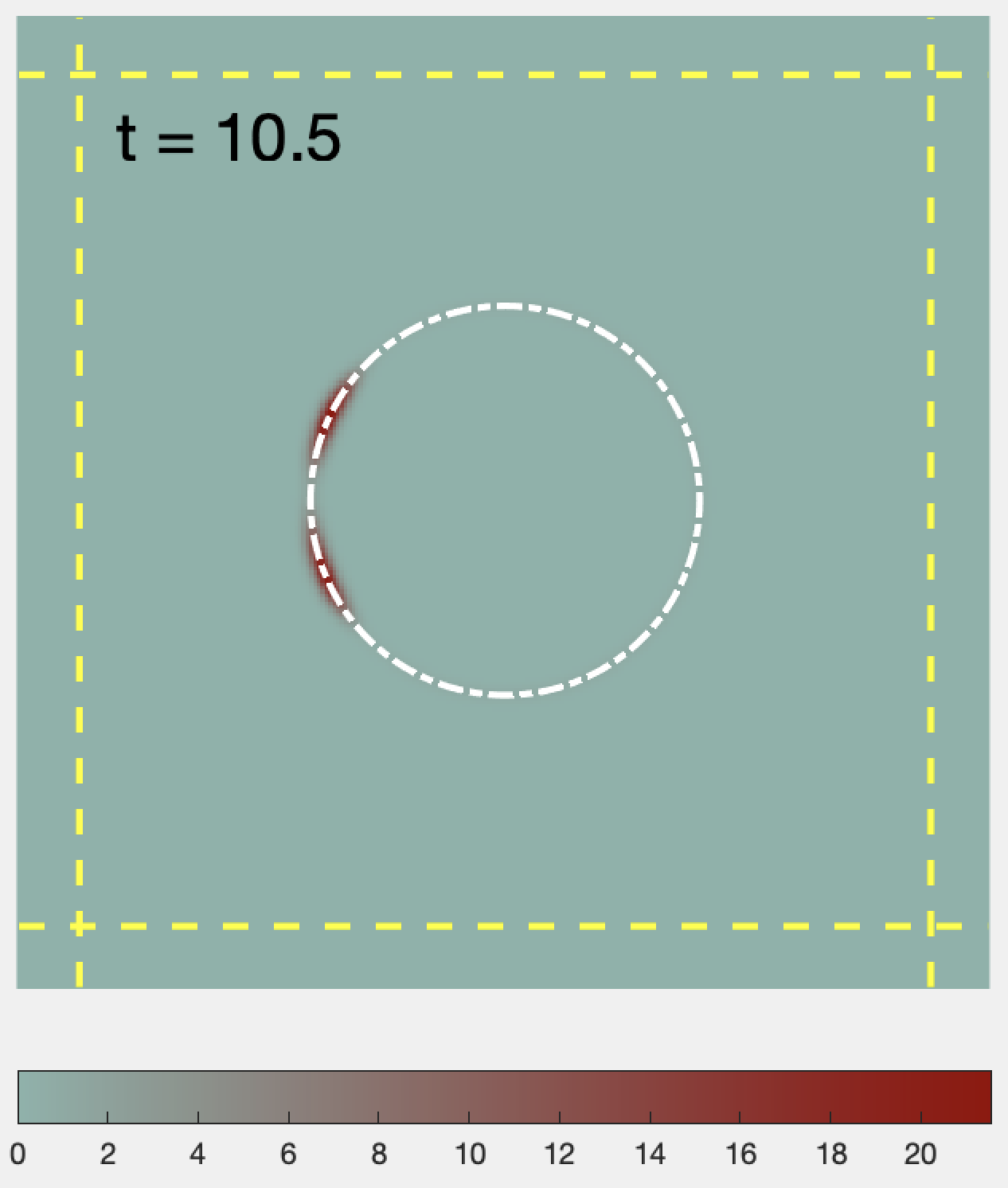}
\includegraphics[width=0.24\linewidth]{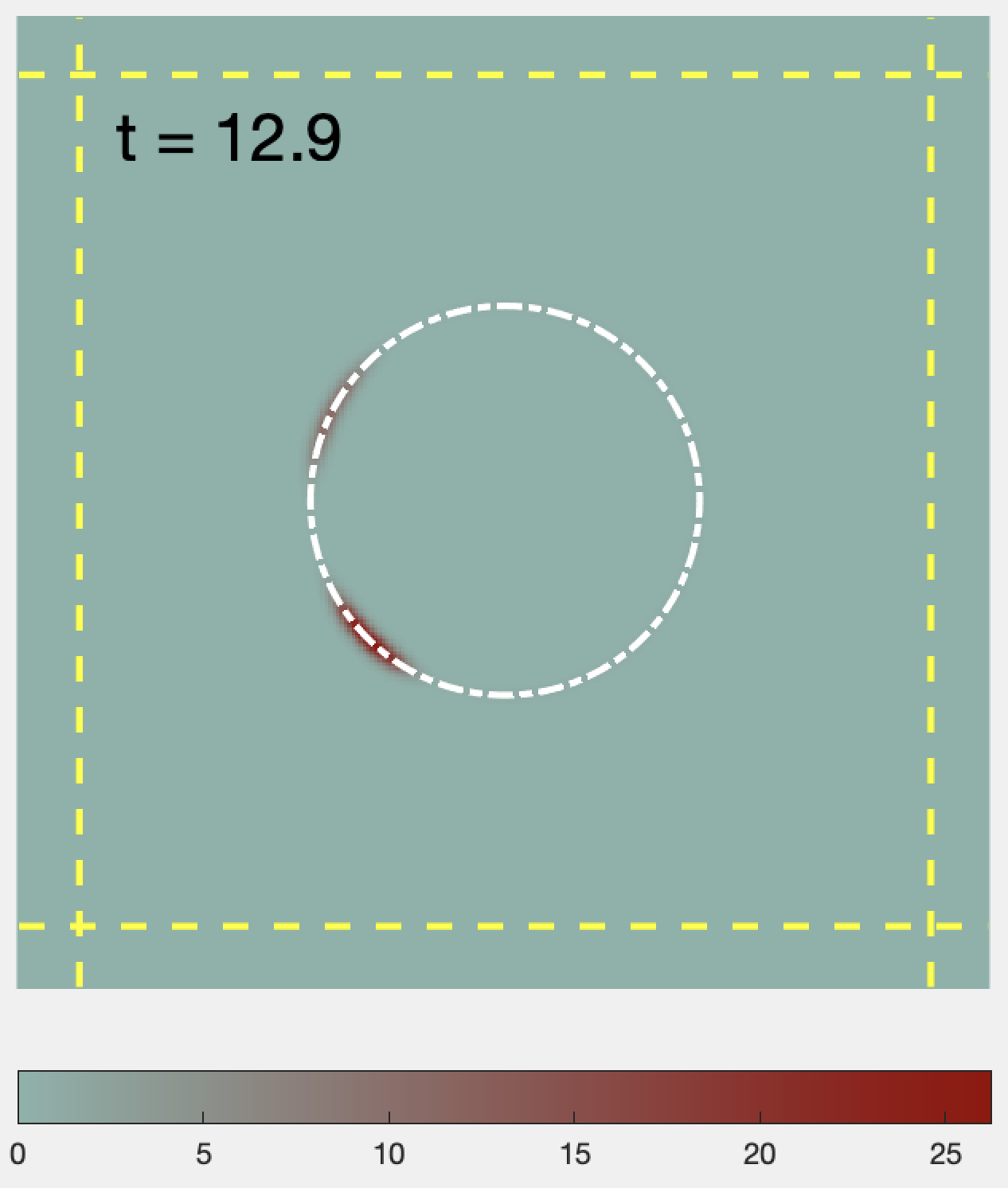} \\
\includegraphics[width=0.24\linewidth]{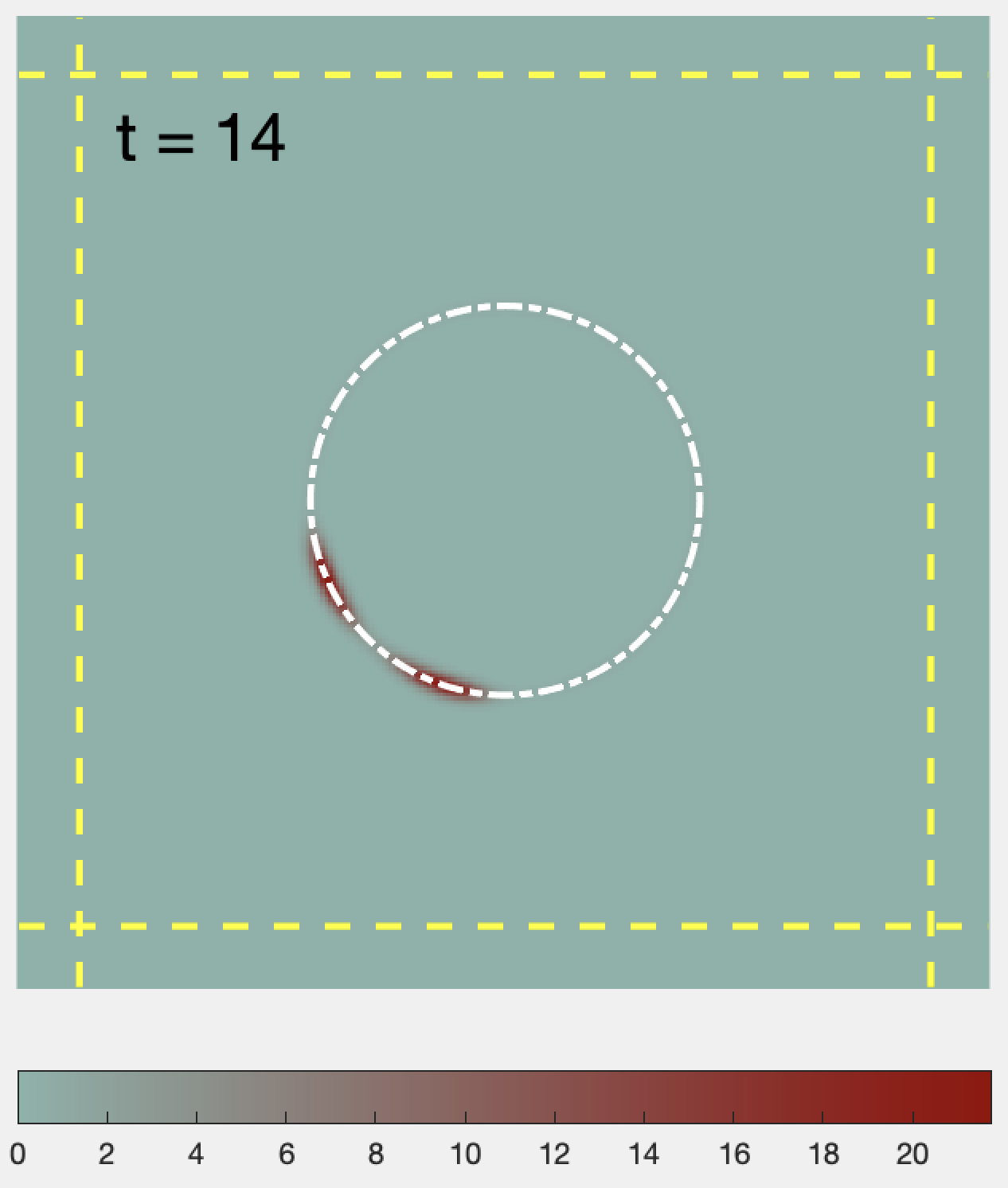}
\includegraphics[width=0.24\linewidth]{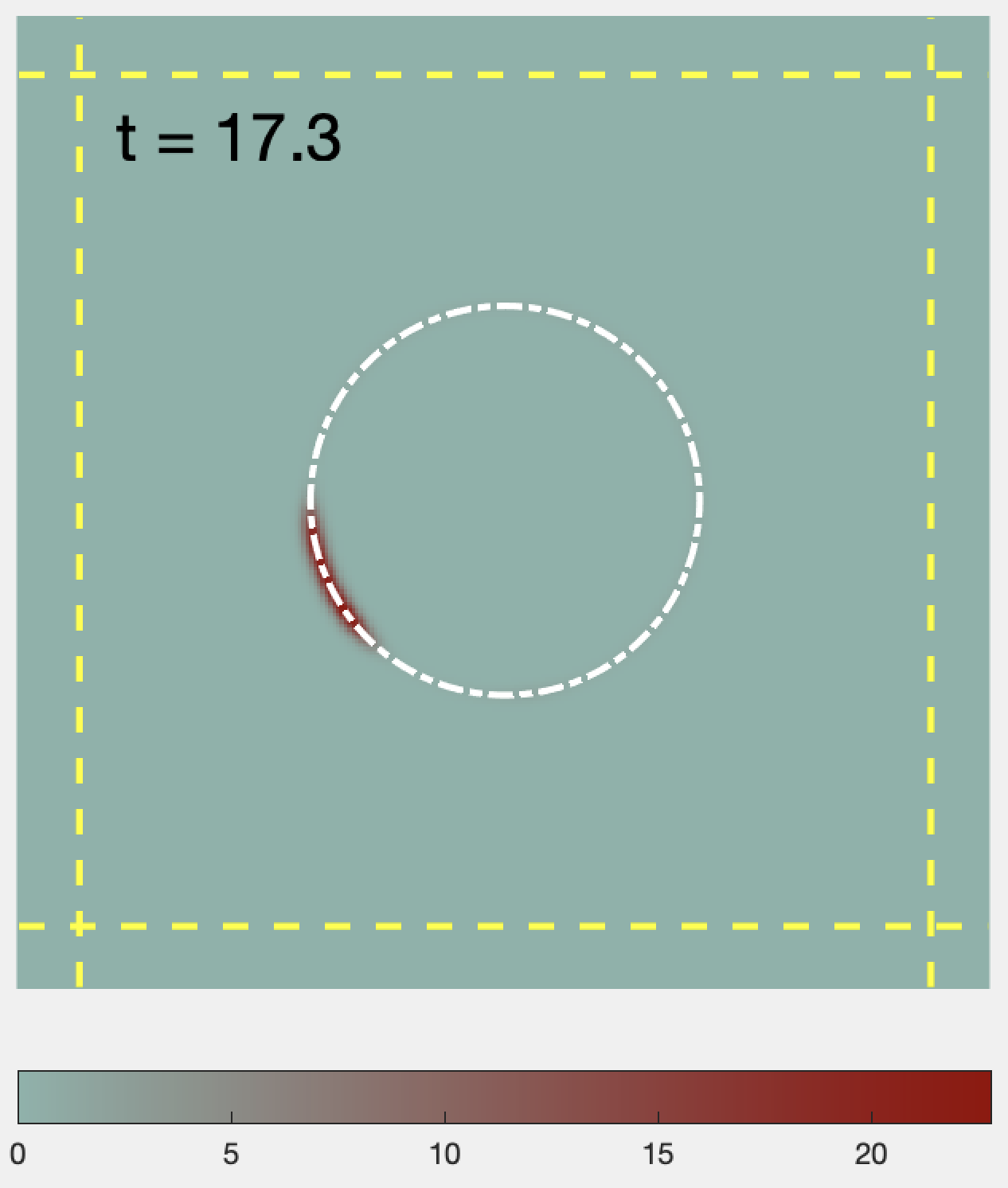}
\includegraphics[width=0.24\linewidth]{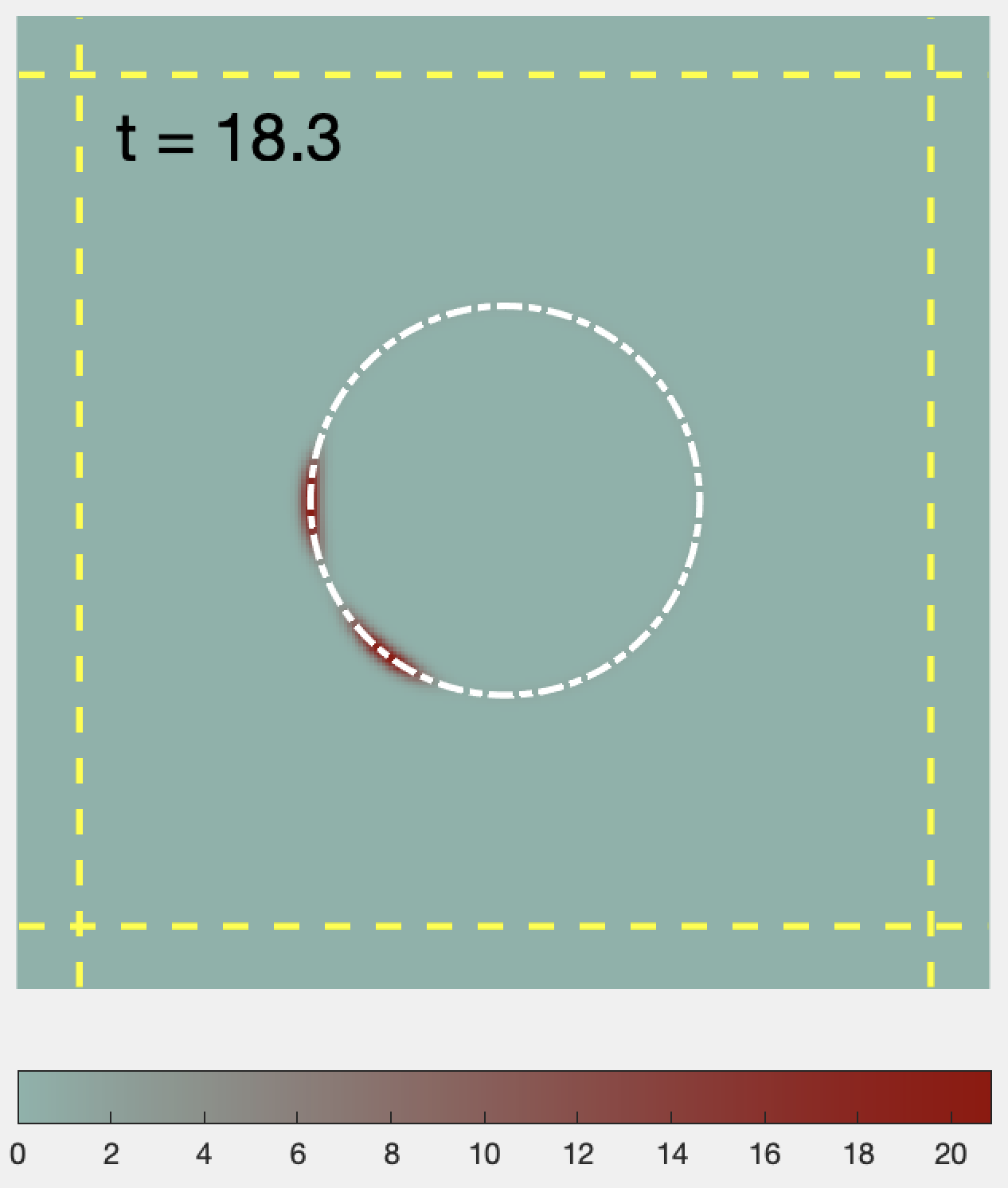}
\end{center}
\caption{ The Meinhardt system coupled with the membrane of a fixed cell: The six subfigures are snapshots at different times. A pattern bifurcation occurs from top left to top middle, followed by competition between the two branches. One branch defeats the other in the top right, and continues to bifurcate into two branches in the bottom left, etc. The white dashed circle indicates the fixed round cell, on which the red indicates the high concentration of the local activator $a$. }
\label{fig:FixedRound} 
\end{figure} 

In this section, we present several numerical experiments of cell chemotaxis dynamics. The numerical method described in the Appendix is adopted to solve the chemotaxis dynamics (\ref{eqn:phi}) coupled with the Meinhardt dynamics (\ref{eqn:a})-(\ref{eqn:c}). In all of the simulations, we take $L_x = L_y = 10, N_x = N_y = 2^8, N_t = 60$ and $\Delta t = 5e-4$. The interfacial width of $\phi$ is fixed as $\epsilon = 10h_x$, where $h_x = 2L_x/N_x$ is the grid spacing in the $x$ direction.

For initial data, we take $\phi^0$ as a disk with center at origin and radius $r = 4$: 
\begin{align}\label{eqn:phi0}
\phi^0(\mathbf{x}) = 0.5 + 0.5\tanh\left( \frac{r - \text{dist}(\mathbf{x},\mathbf{0})}{\epsilon/3} \right),
\end{align}
in which $\text{dist}(\mathbf{x},\mathbf{0})$ stands for the Euclidean distance between $\mathbf{x}$ and the origin. We further take 
\[
a^0(\mathbf{x}) \equiv 0, \ b^0(\mathbf{x}) \equiv 0.01, \ c^0(\mathbf{x}) \equiv 0.
\]
Unless otherwise specified, the chemoattractant source is located at $\mathbf{r}_0 = (0, -40)^T$ and the strength of the bias  is $C_{\text{chem}} = 0.02$.

\subsection{Meinhardt dynamics on the membrane of a fixed cell}

In this example, we test the Meinhardt dynamics on the membrane of a fixed phase field cell as given in (\ref{eqn:phi0}). The numerical simulation is presented in Figure \ref{fig:FixedRound} from which the bifurcating patterns are clearly observed on the cell membrane. In this simulation, several time snapshots are taken at $t = 10.0, 11.0, 13.0, 14.0, 16.8, 18.0$. A bifurcation occurs at $t = 10.0$. After a short time period, two branches are formed at $t = 11.0$. One branch defeats the other at $t = 13.0$. Then the bifurcation repeatedly recurs. In each subfigure, the black dashed circle represents the cell membrane. The yellow dashed lines indicate the small box in which the Meinhardt equations are solved (see Appendix for the details about the small box).

\begin{figure}[t] 
\begin{center}
\includegraphics[width=0.24\linewidth]{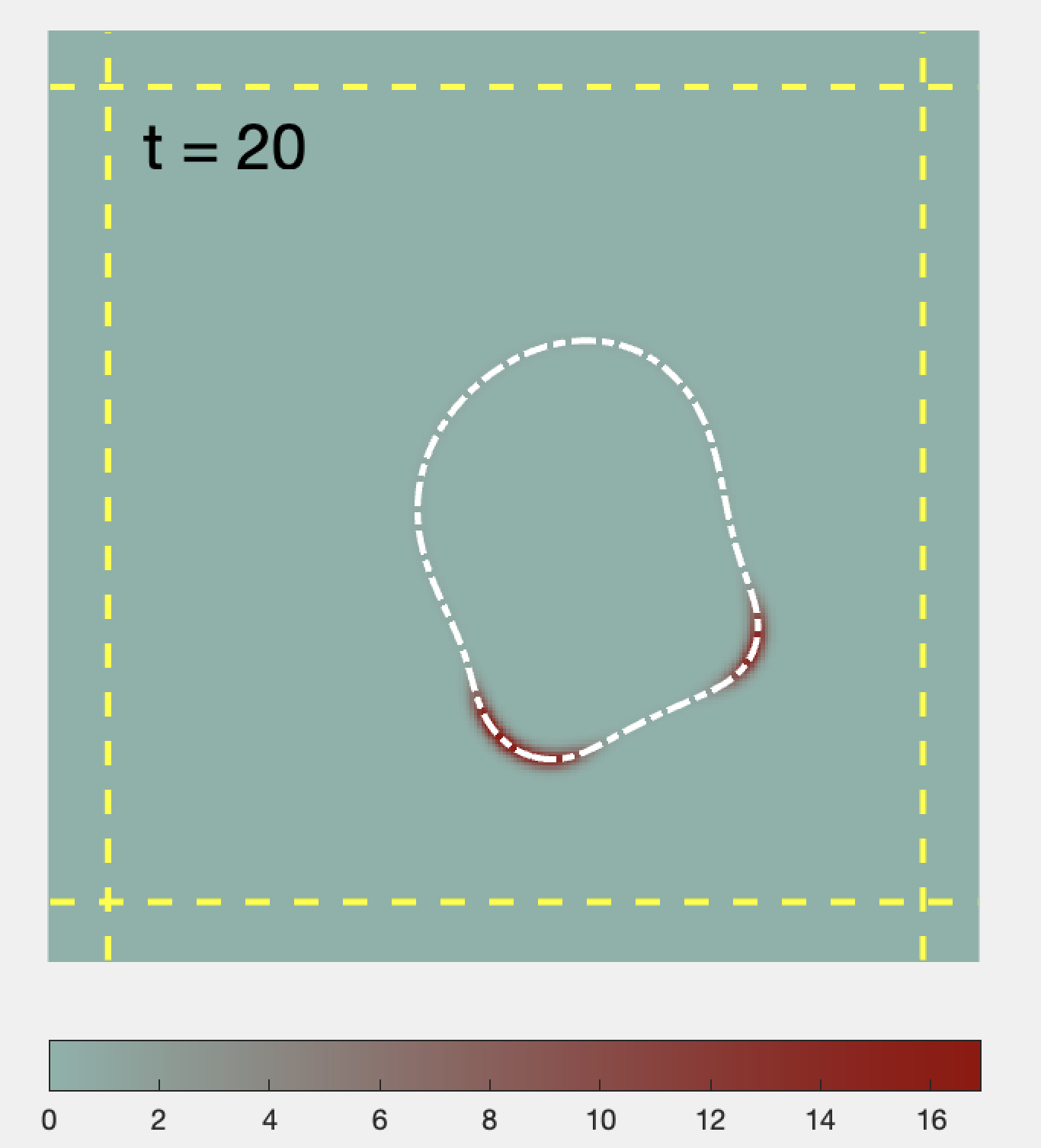}
\includegraphics[width=0.24\linewidth]{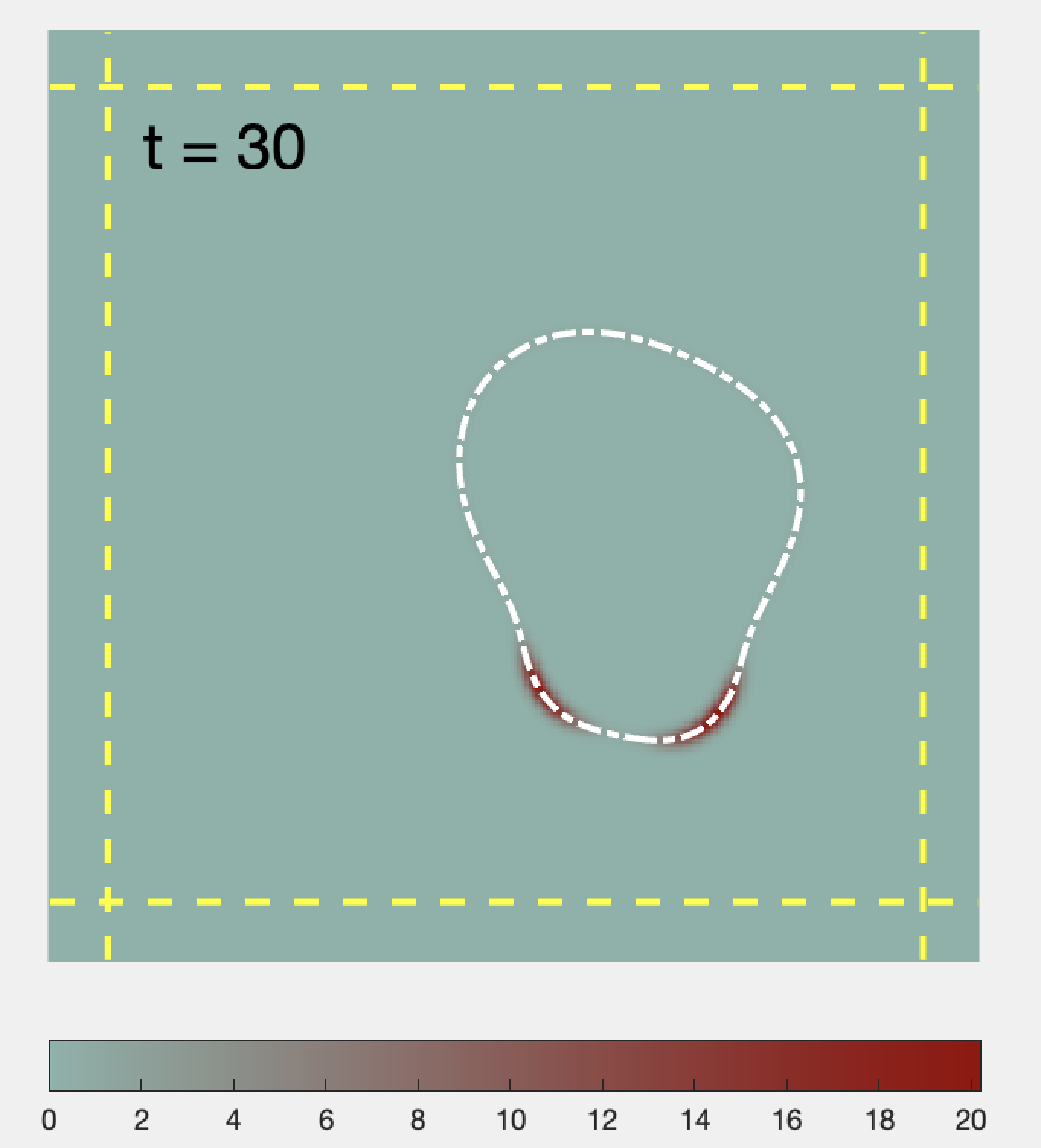}
\includegraphics[width=0.24\linewidth]{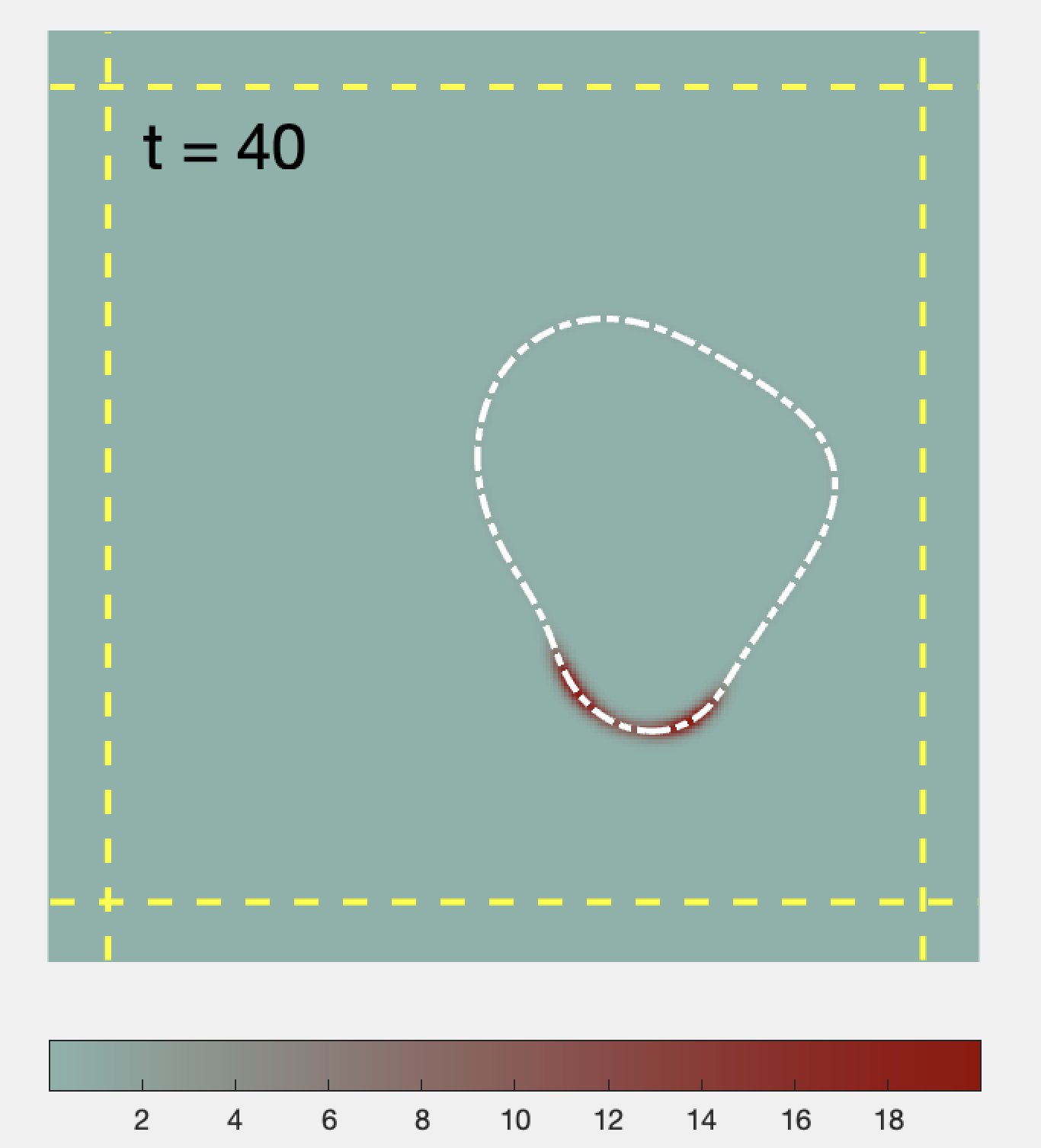} 
\includegraphics[width=0.24\linewidth]{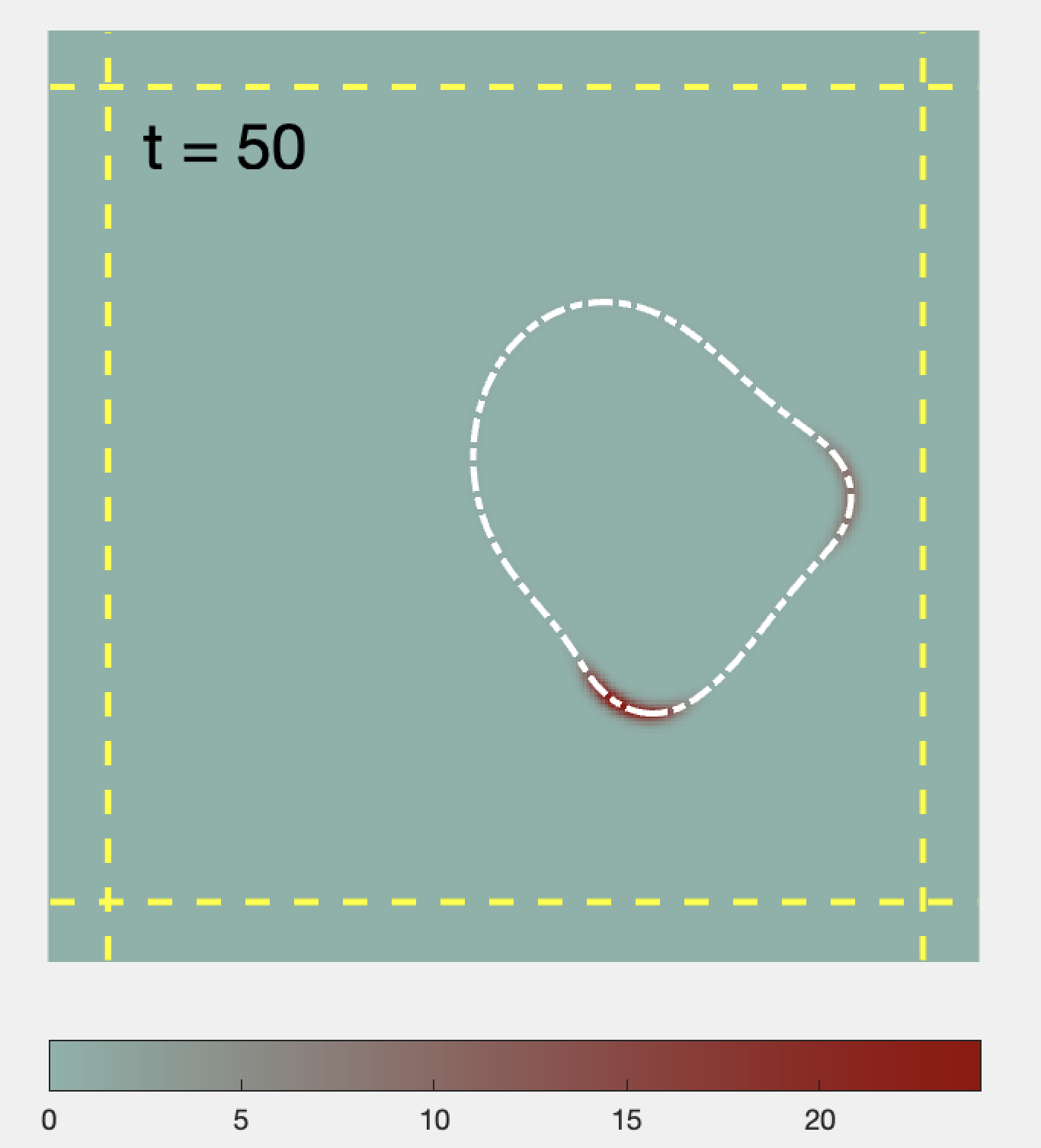}\\
\includegraphics[width=0.24\linewidth]{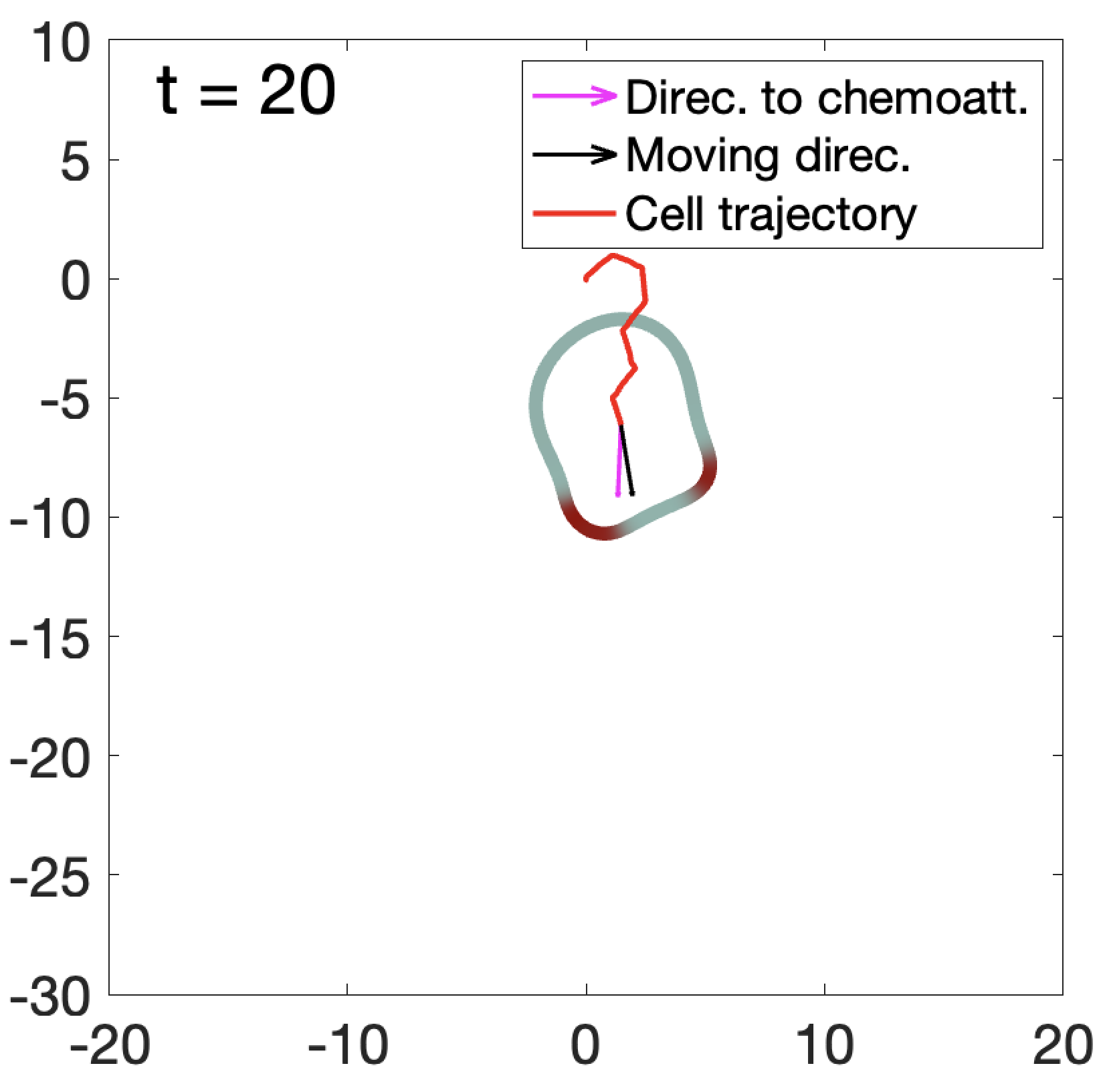}
\includegraphics[width=0.24\linewidth]{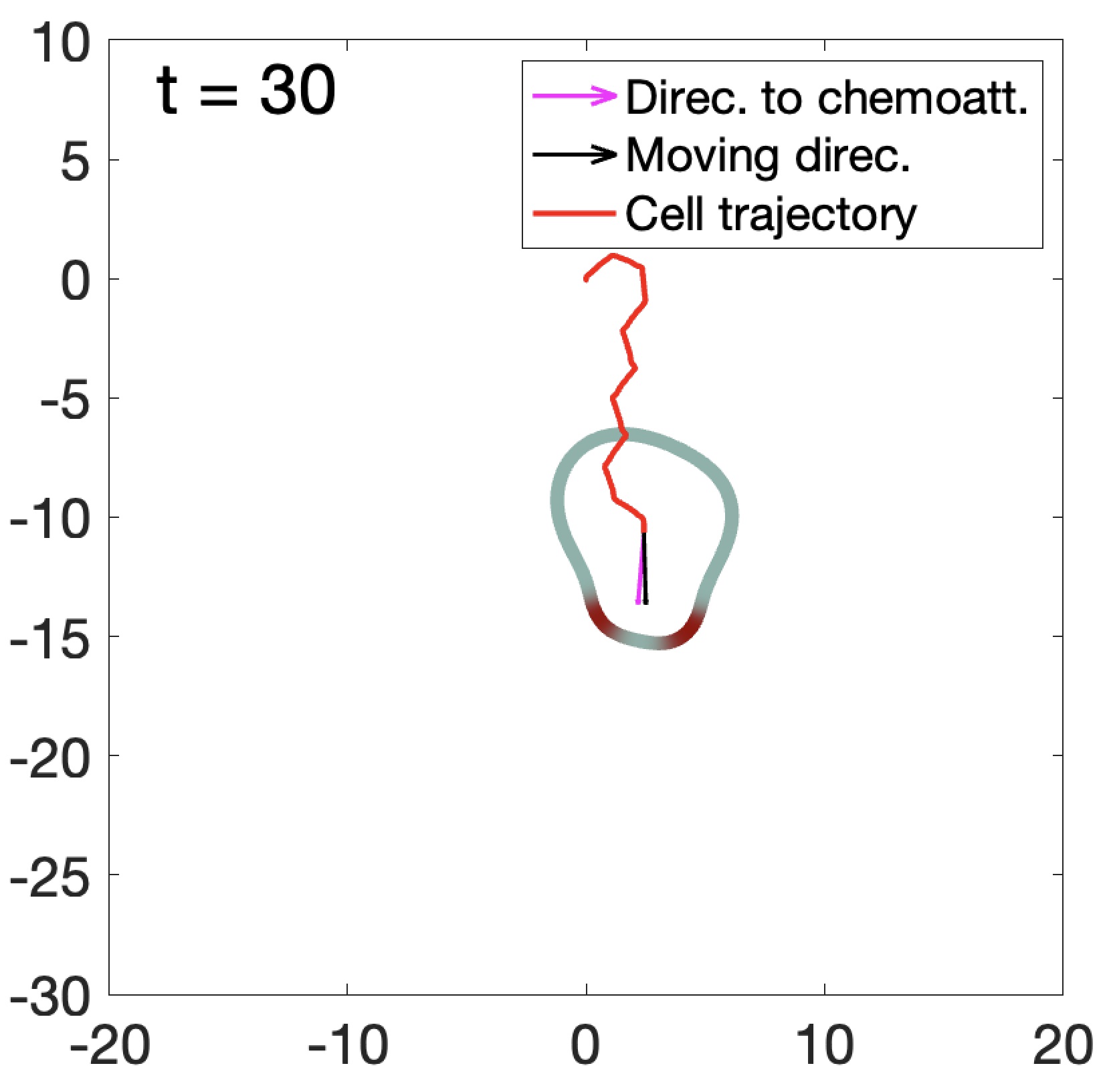}
\includegraphics[width=0.24\linewidth]{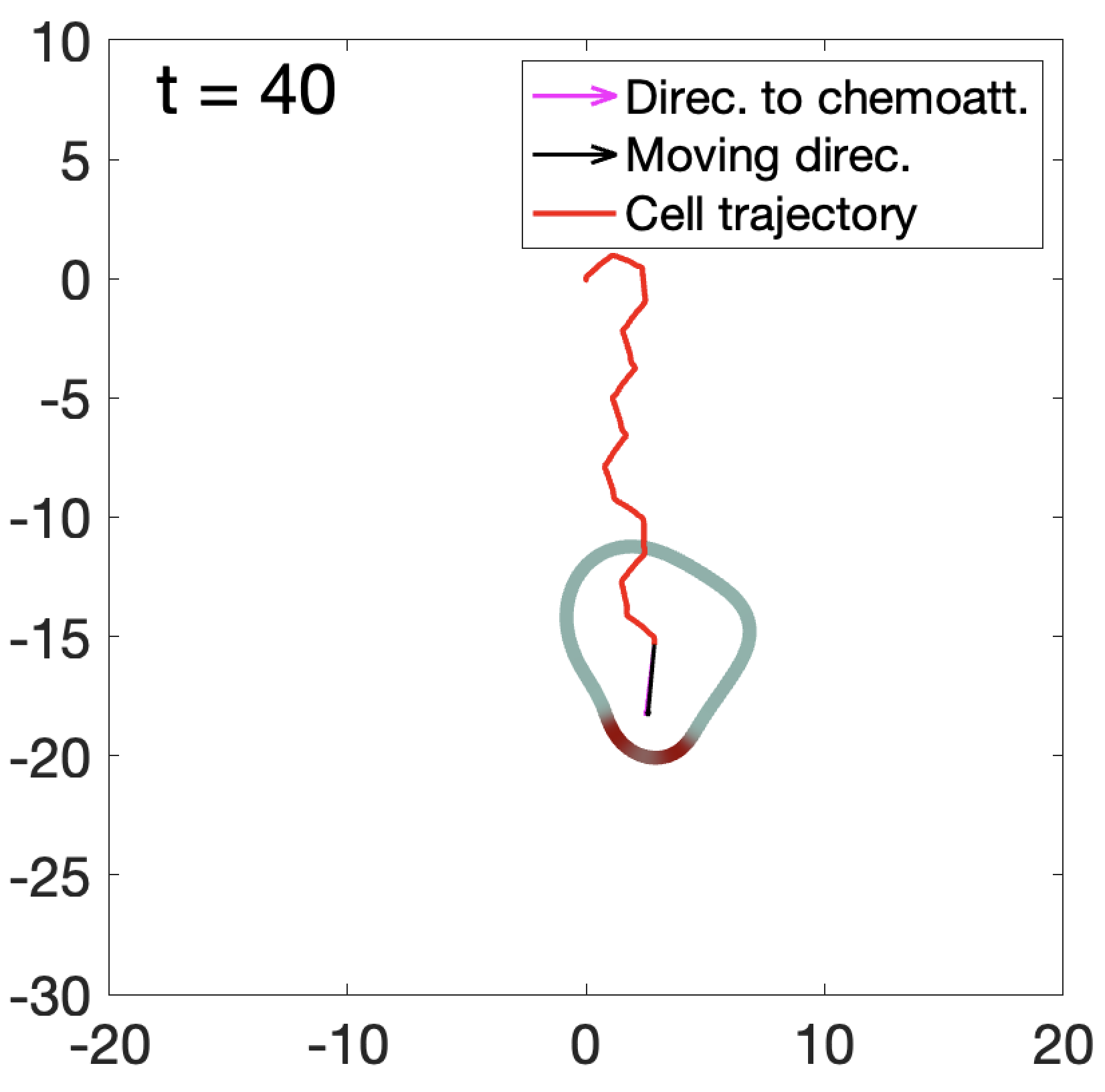} 
\includegraphics[width=0.24\linewidth]{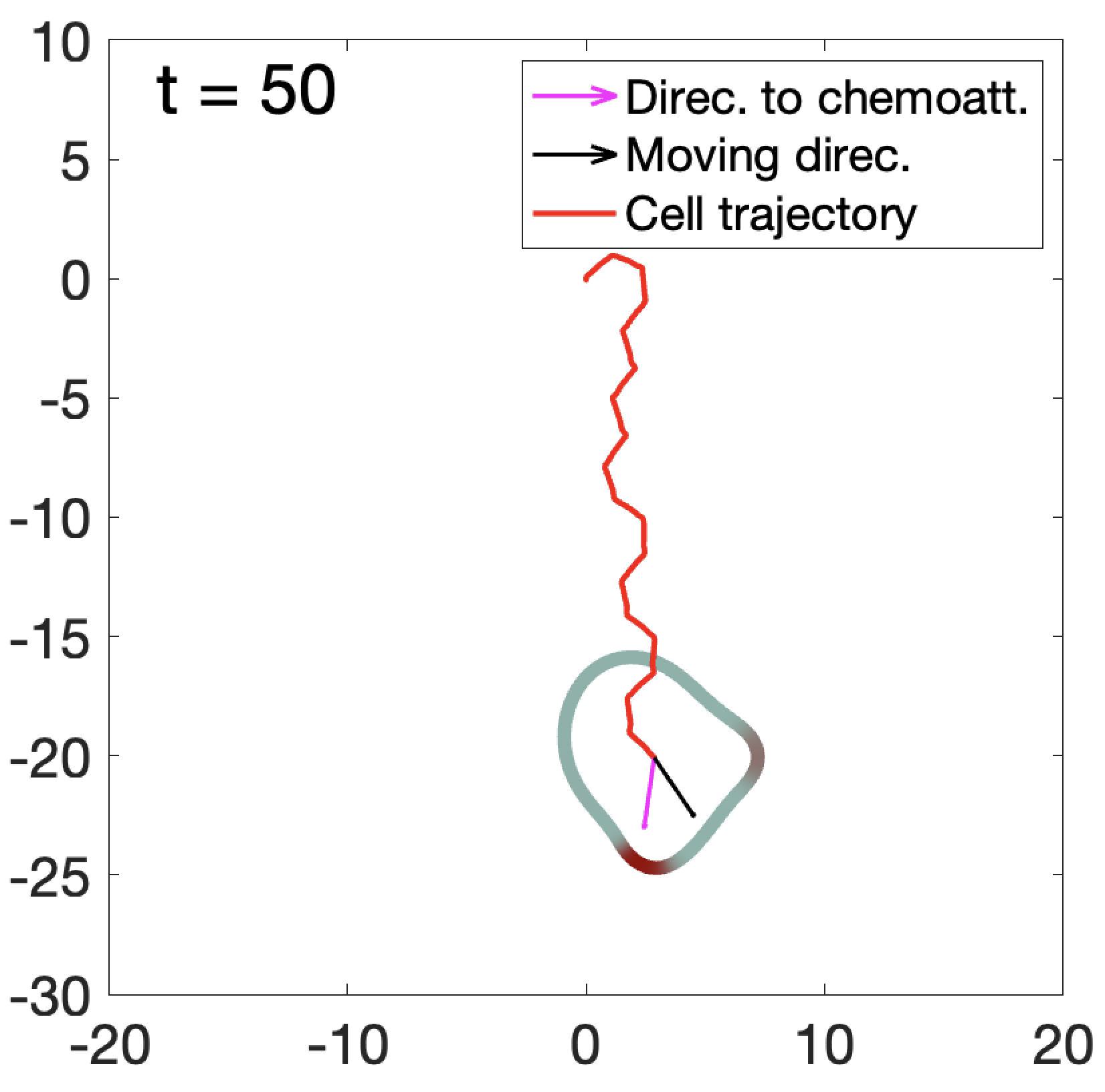}\\
\end{center}
\caption{Top row: The phase field model of cell movement driven by the Meinhardt reaction-diffusion process on the membrane. The four subplots are snapshots at $t=20,30,40,50$ respectively. In each subplot, the cell membrane is visualized by the 1/2-level set of $\phi$ (the white dashed curve), on which the red color indicates high concentration of activator $a$ in the Meinhardt model. Bottom row: trajectory of the simulated cell towards chemoattractant source $\textbf{r}_0$ ($\textbf{r}_0 = [0,-40]^T$, which is not shown in the subplots) at time $t=20,30,40,50$. The cell curve is colored by a colormap to indicate the concentration of activator $a$. The red curve represents the trajectory of the simulated cell, while the purple and black arrows respectively represent the direction of source and cell's center-of-mass velocity. }
\label{fig:Mov01} 
\end{figure} 

\subsection{Meinhardt dynamics on the membrane of a free cell}

We now allow the cell to deform and move by set the cell solving the phase field equation (\ref{eqn:phi}) together with the Meinhardt system (\ref{eqn:a})-(\ref{eqn:c}). A set of numerical results is presented in Figure \ref{fig:Mov01}. In this figure, the top row is the four snapshots at $t = 20,30,40,50$, on which the yellow color indicates a high concentration of activator $a$. The bottom row shows the cell trajectories for times up to $t = 20,30,40,50$. Our simulations turn out to be generally consistent with observed chemotactic behavior. Pseudopods are randomly generated in the cell membrane, while the biasing effects towards the direction of the chemoattractant source $\textbf{r}_0$  accumulate over time, which eventually leads to the cell’s translation along favorable directions. The trajectory of our simulated cell has demonstrated high efficiency in the cell’s navigation. 

\begin{figure}[t] 
\begin{center}
\includegraphics[width=0.24\linewidth]{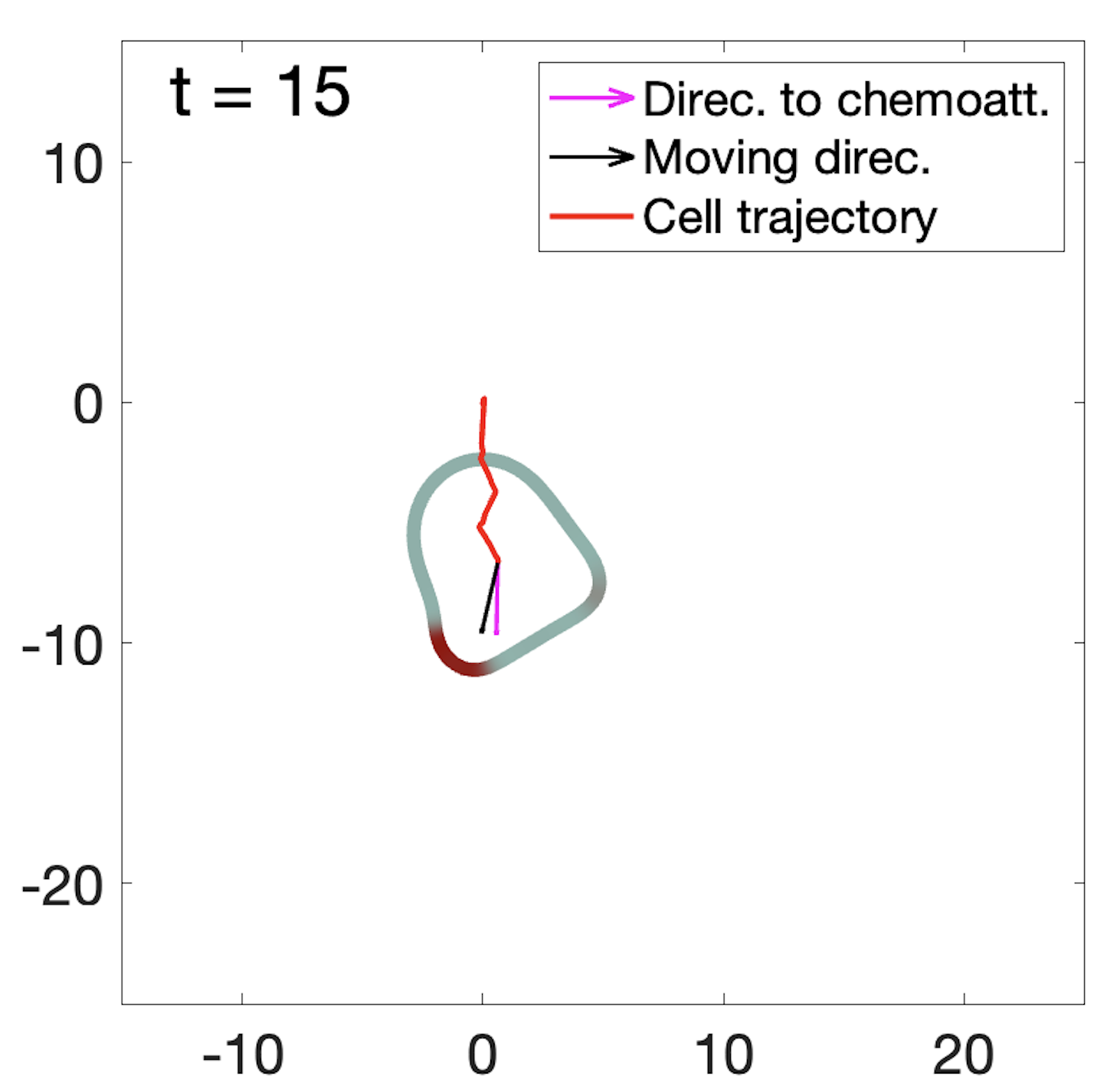}
\includegraphics[width=0.24\linewidth]{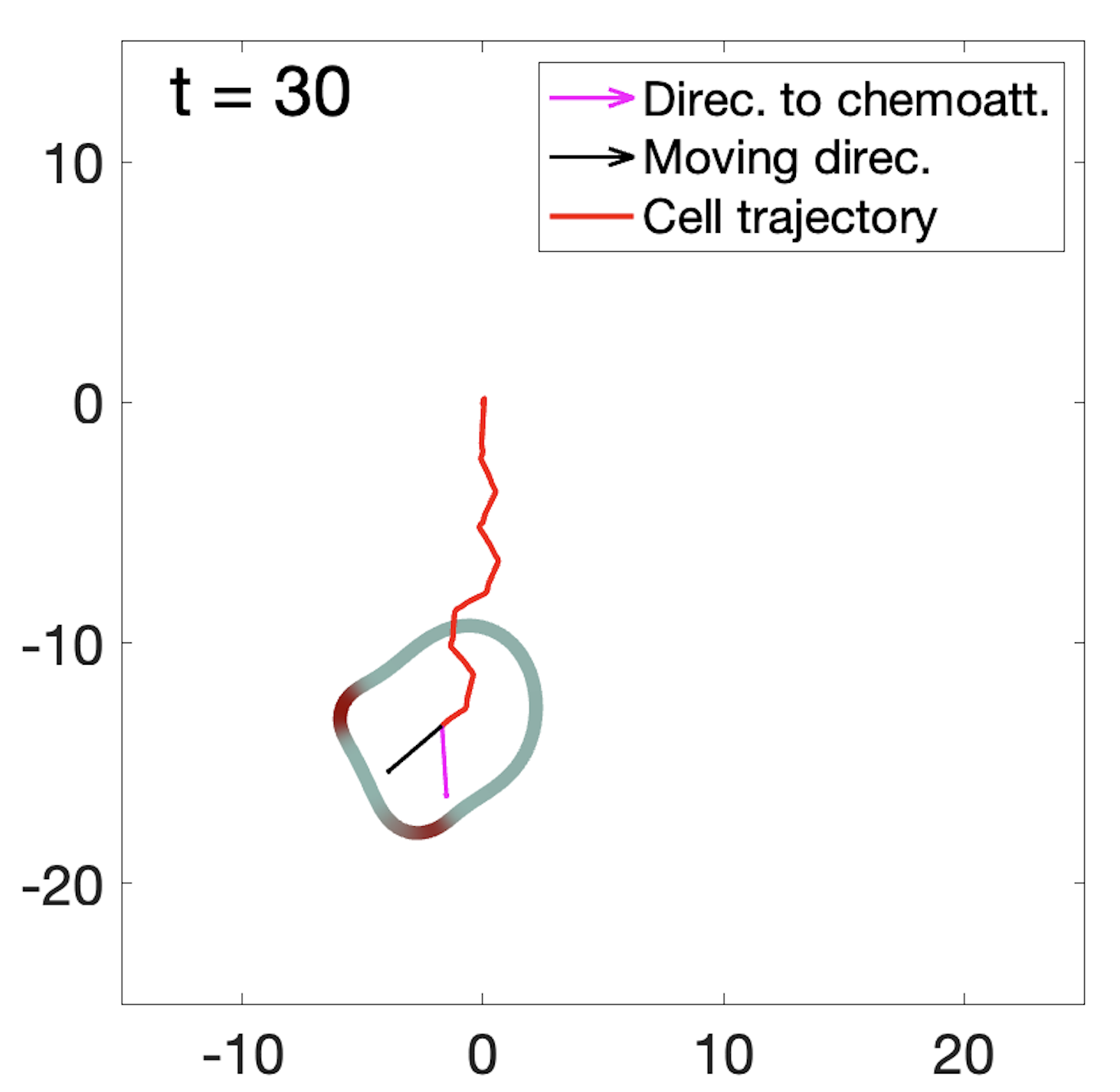}
\includegraphics[width=0.24\linewidth]{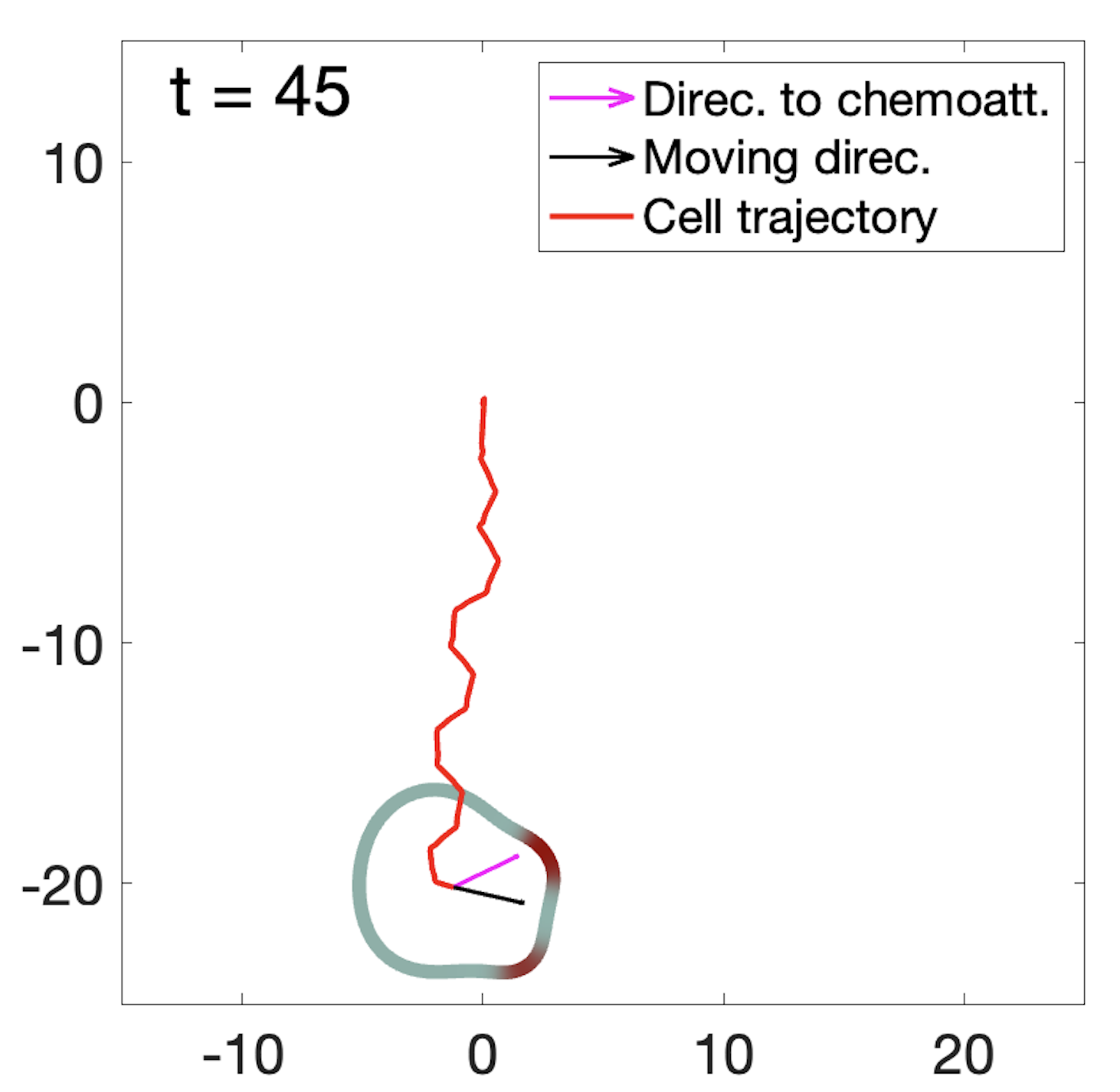} 
\includegraphics[width=0.24\linewidth]{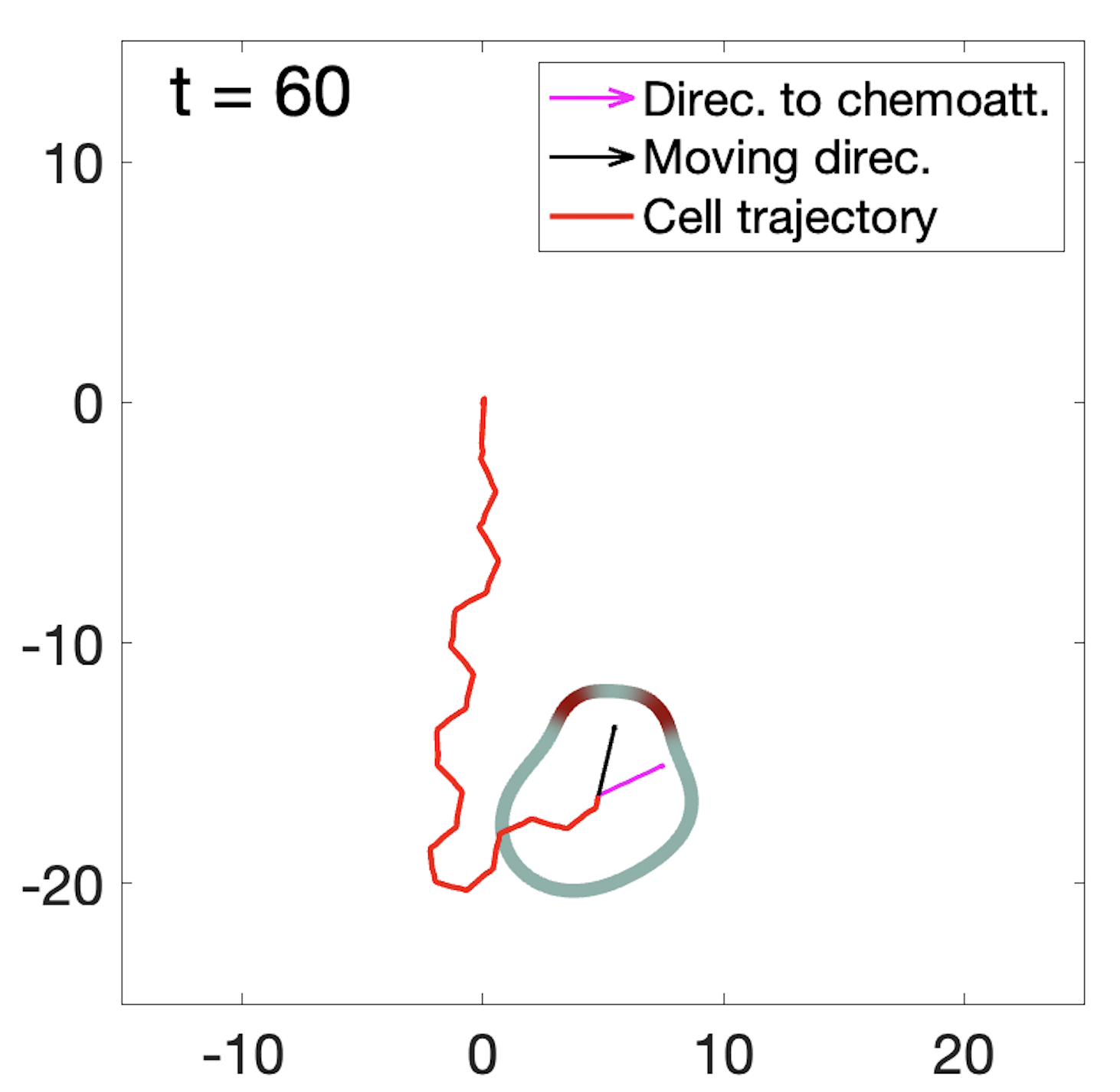}\\
\includegraphics[width=0.24\linewidth]{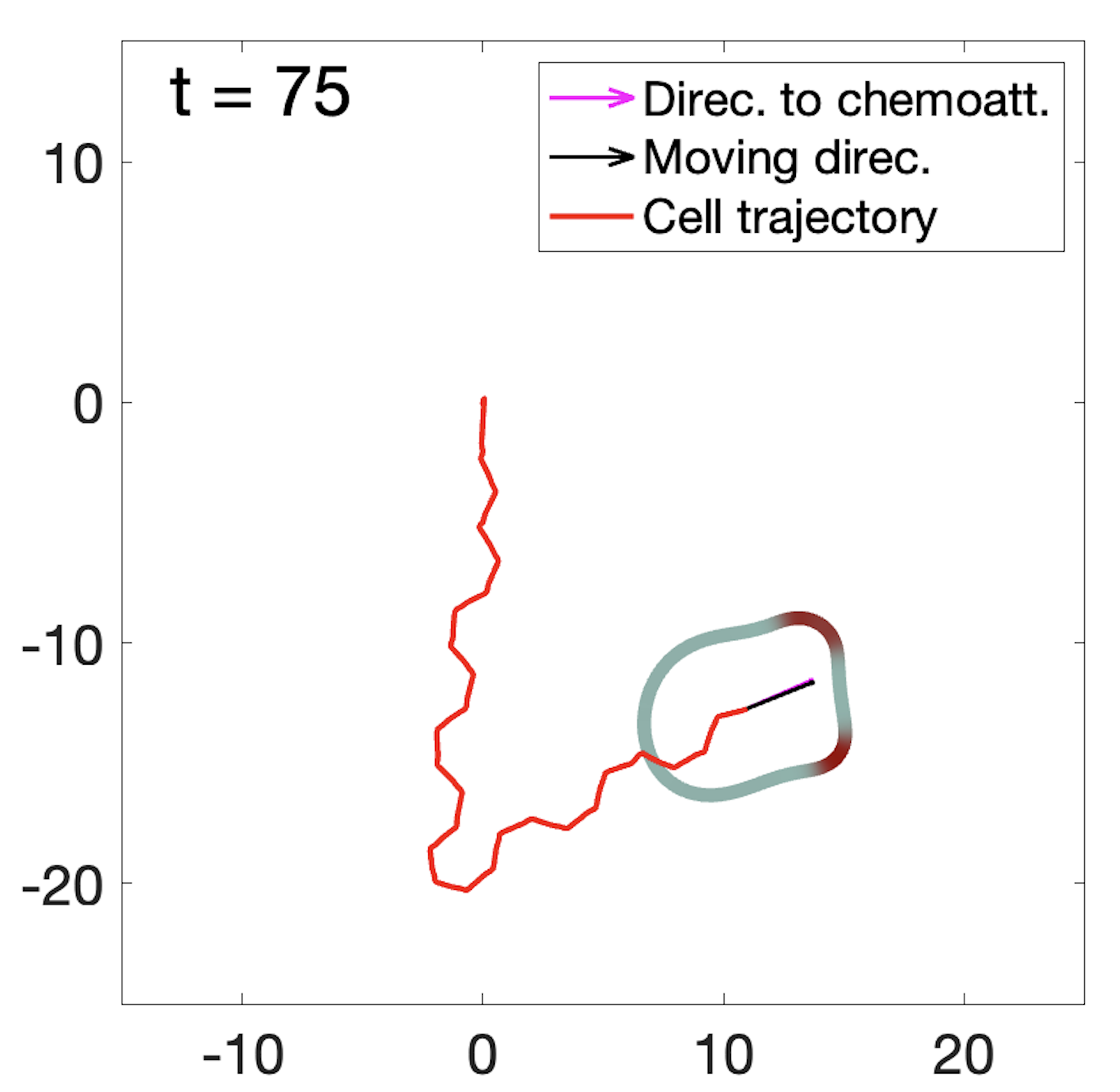}
\includegraphics[width=0.24\linewidth]{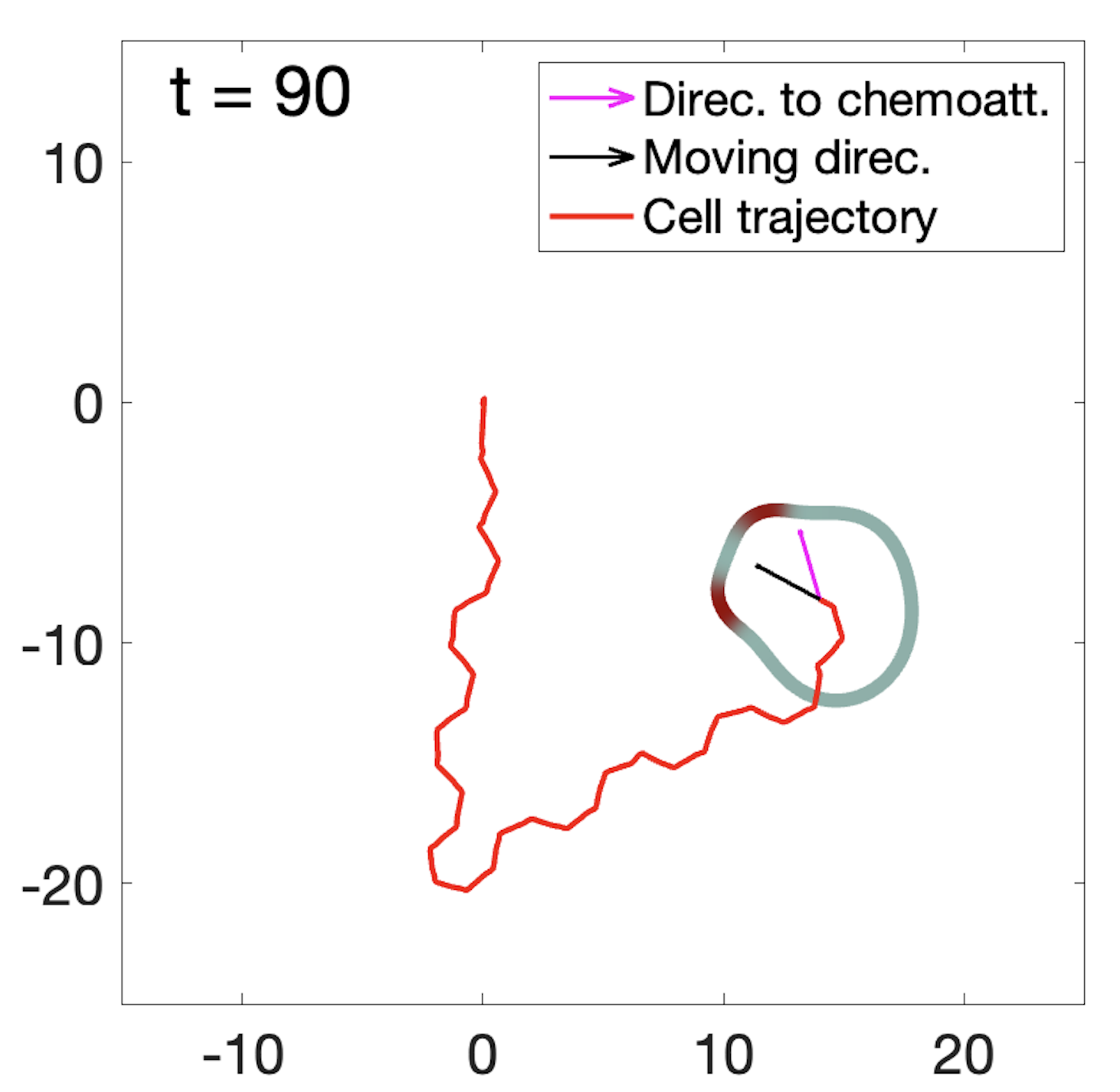}
\includegraphics[width=0.24\linewidth]{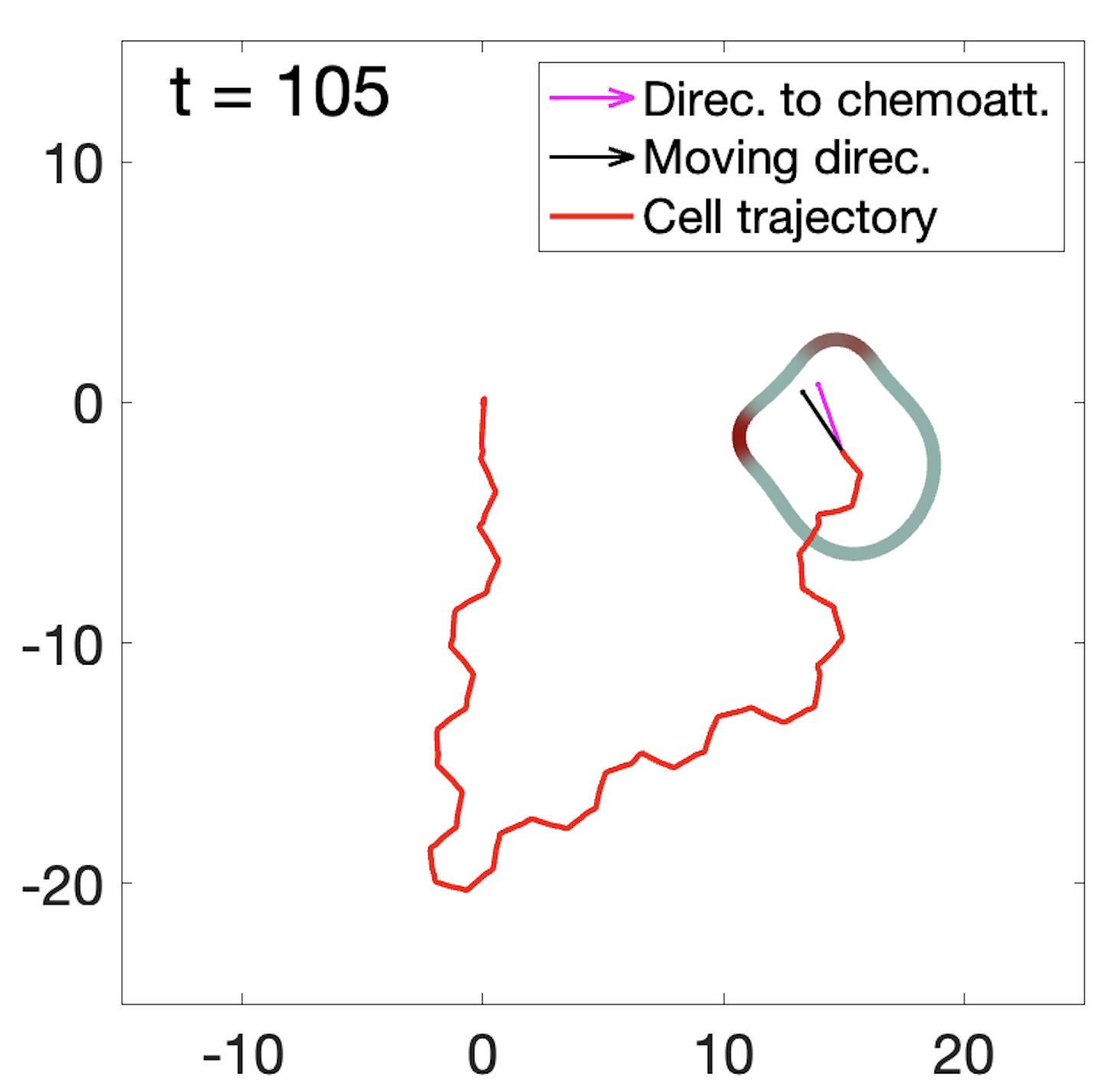} 
\includegraphics[width=0.24\linewidth]{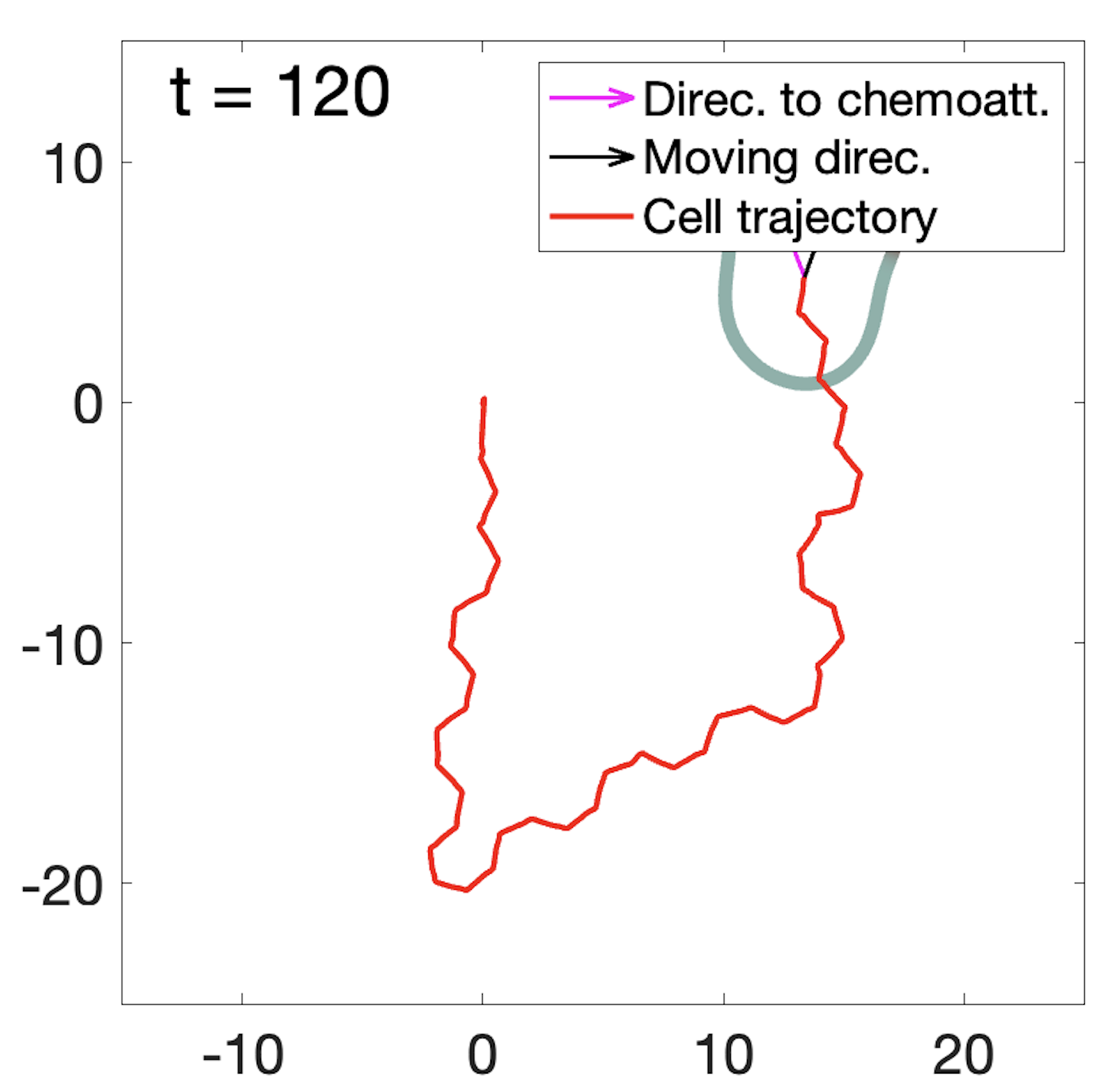}\\
\end{center}
\caption{Trajectory of the simulated cell when the source $\mathbf{r}_0$ suddenly relocates. All conditions are the same as those in Figure \ref{fig:Mov01}, except that the chemoattractant source is changed to a new location every 40s. }
\label{fig:Mov02} 
\end{figure}

Moreover, even when the chemoattractant source $\mathbf{r}_0$ is suddenly moved, a strong adaptability of our simulated cell is also observed in Figure \ref{fig:Mov02}. In this simulation, the chemoattractant source $\mathbf{r}_0$ is located at $\mathbf{r}_0 = (0, -40)^T$ over time $[0,40]$. During this time period, the cell moves towards $\mathbf{r}_0$ similarly to that shown in Figure \ref{fig:Mov01}. At $t = 40$, $\mathbf{r}_0$ is suddenly changed to $\mathbf{r}_0 = (40, 0)^T$. The cell can quickly adjust the direction and move towards the new location over time $[40,80]$, still with a slight biasing effect. At $t = 80$, a new location $\mathbf{r}_0 = (0,40)^T$ is assigned, and the cell changes direction and moves towards the north. For the sake of clear display, the cells are plotted in a window of $[-15,25]\times[-25,15]$, without showing the locations of chemoattractant source.

\subsection{Chemotaxis index}

To quantify the efficiency of cells' navigation, we measured the chemotaxis index (CI) of our simulated cells over time. The chemotaxis index is defined by the ratio between the distance the simulated cell has traveled in the direction toward the chemoattractant source $\mathbf{r}_0$ and the total distance it has traveled:
\[
\text{CI} = \frac{ \sum_n \Big\langle \mathbf{x}_{\text{center}}^{n+1}-\mathbf{x}_{\text{center}}^{n}, \frac{\mathbf{r}_0 - \mathbf{x}_{\text{center}}^{n} } {\|\mathbf{r}_0 - \mathbf{x}_{\text{center}}^{n}\|} \Big\rangle  }{ \sum_n  \Big\langle \mathbf{x}_{\text{center}}^{n+1}-\mathbf{x}_{\text{center}}^{n},  \mathbf{x}_{\text{center}}^{n+1}  - \mathbf{x}_{\text{center}}^{n} \Big\rangle }.
\]

\begin{figure}[t] 
\begin{center}
\includegraphics[width=0.3\linewidth]{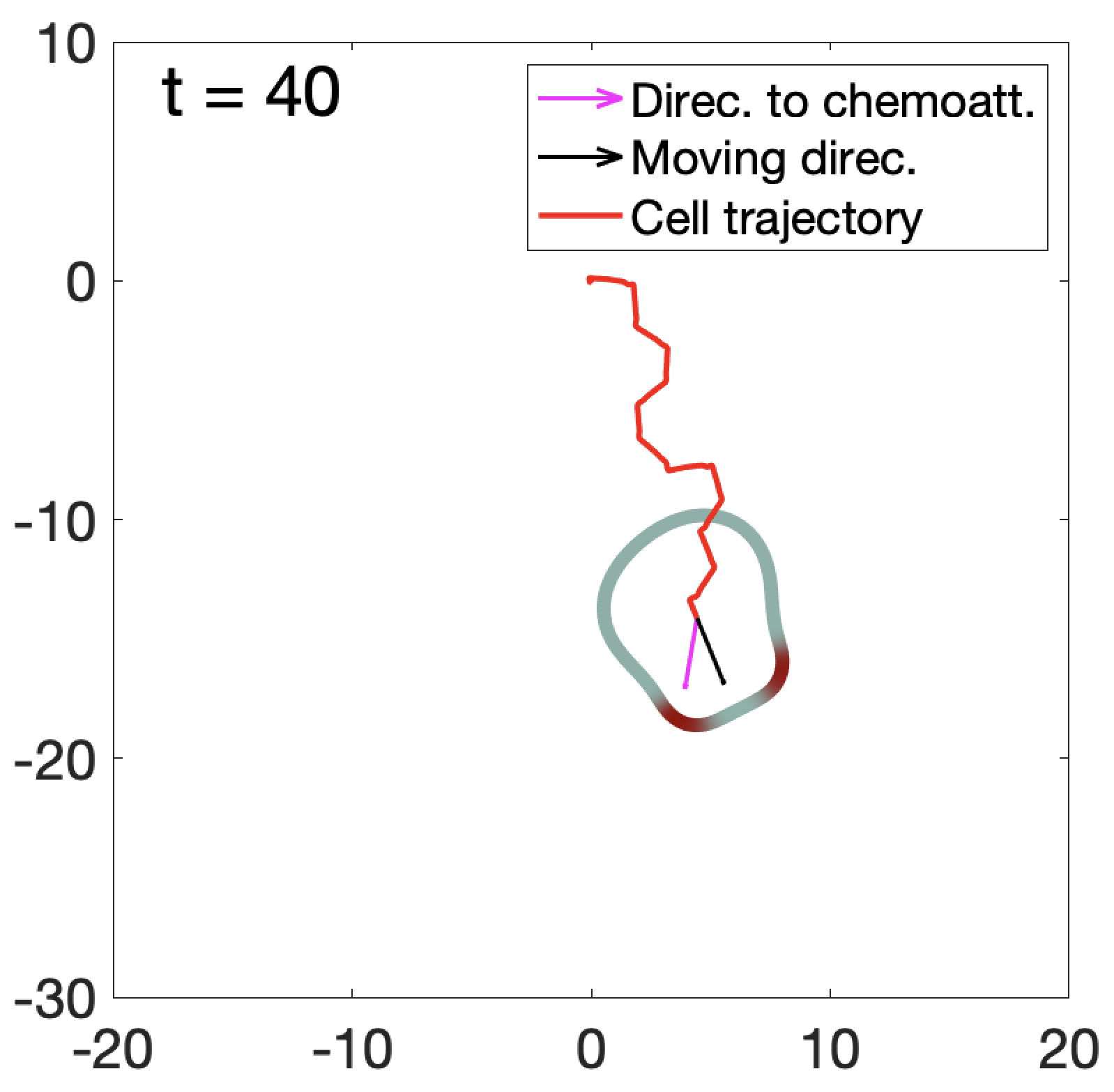}
\includegraphics[width=0.3\linewidth]{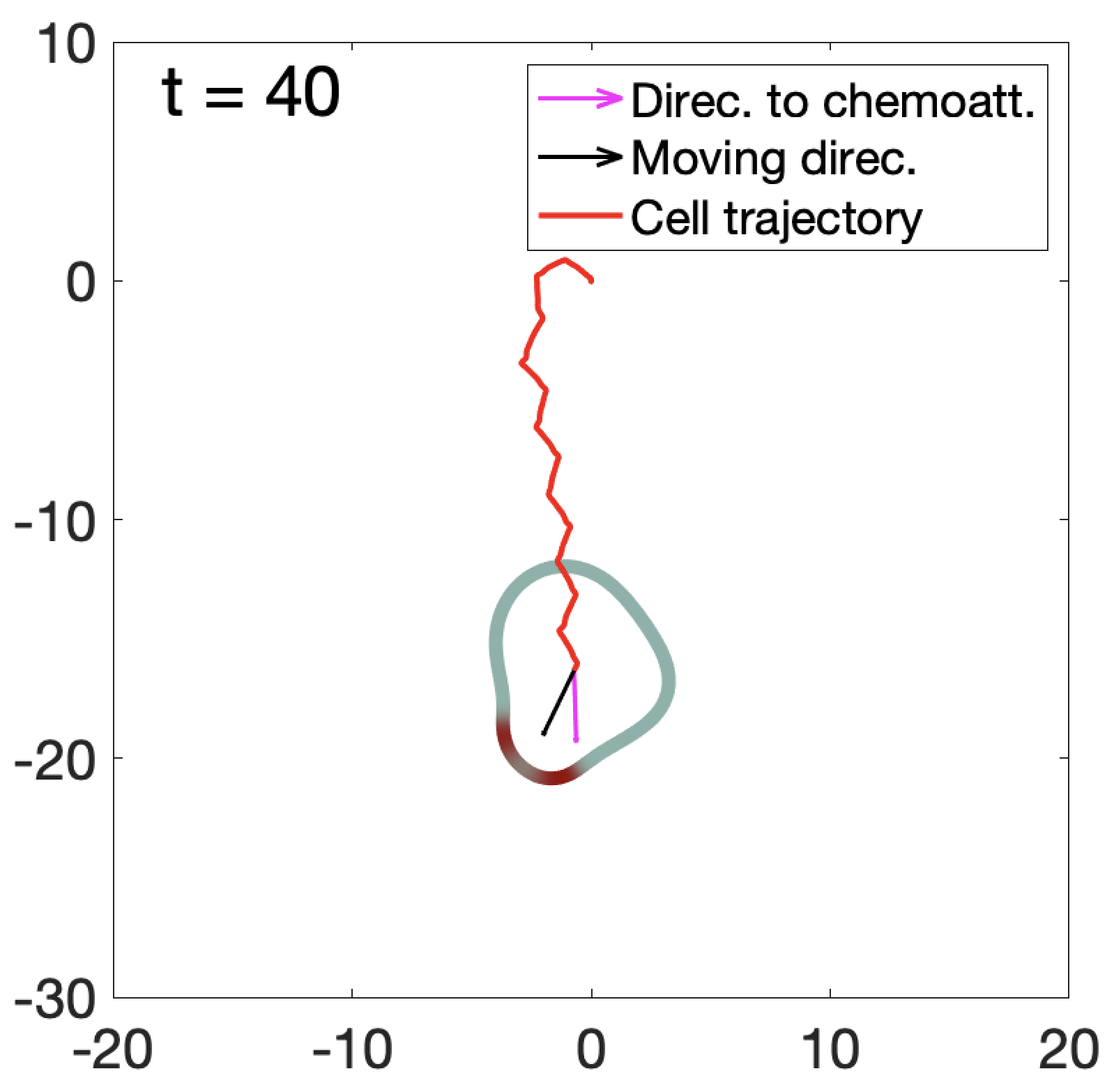}
\includegraphics[width=0.3\linewidth]{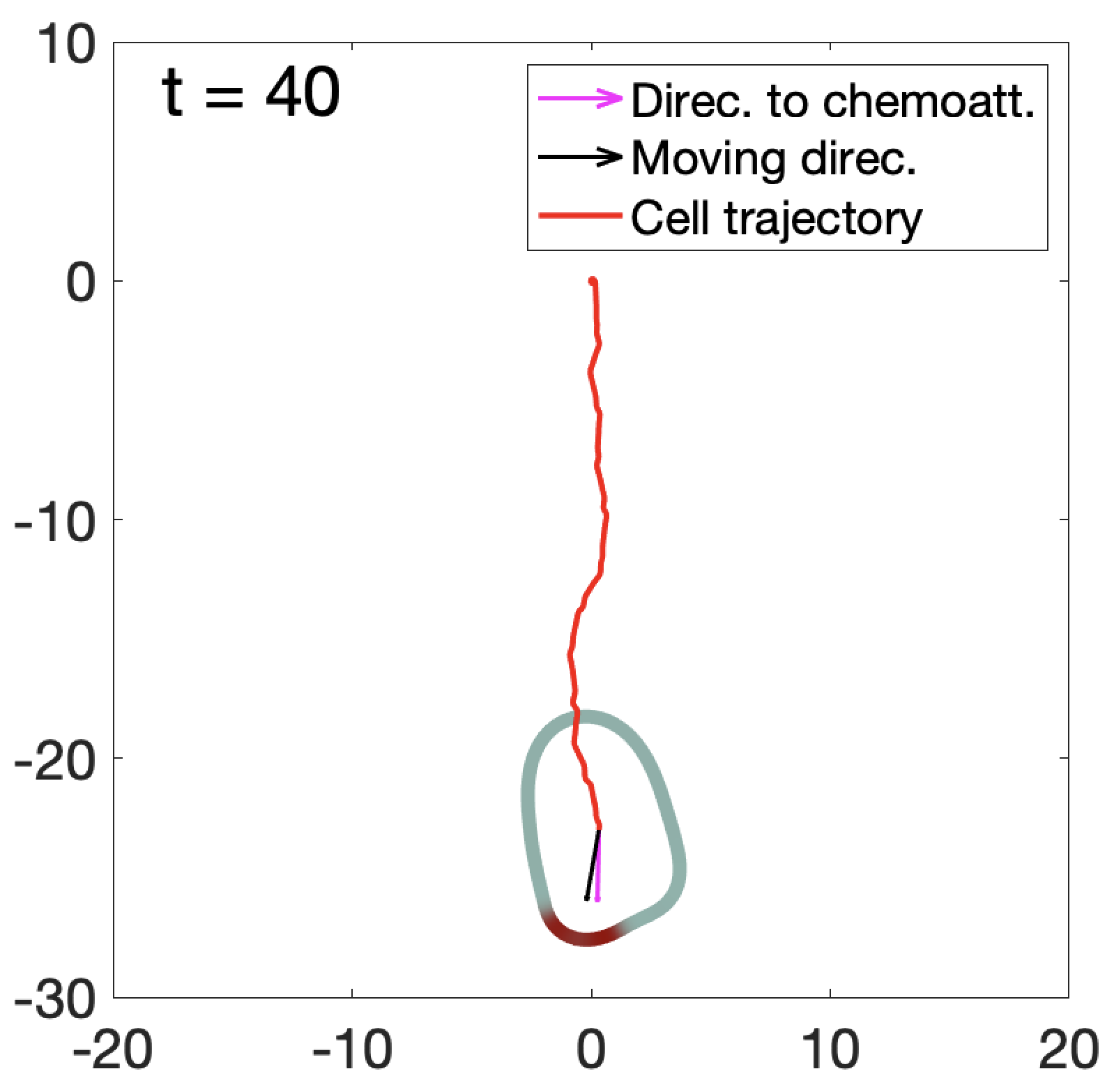} \\
\includegraphics[width=0.3\linewidth]{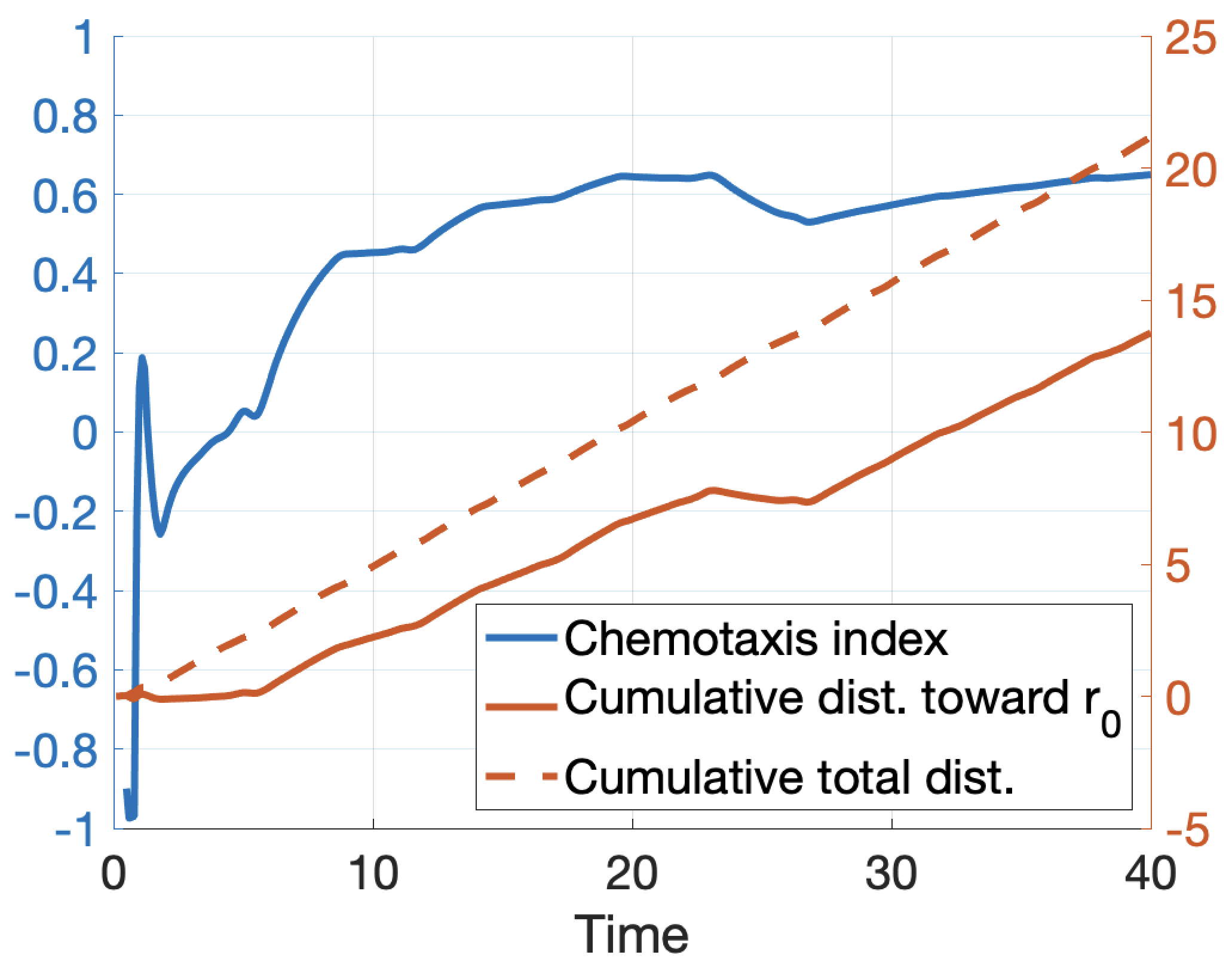}
\includegraphics[width=0.3\linewidth]{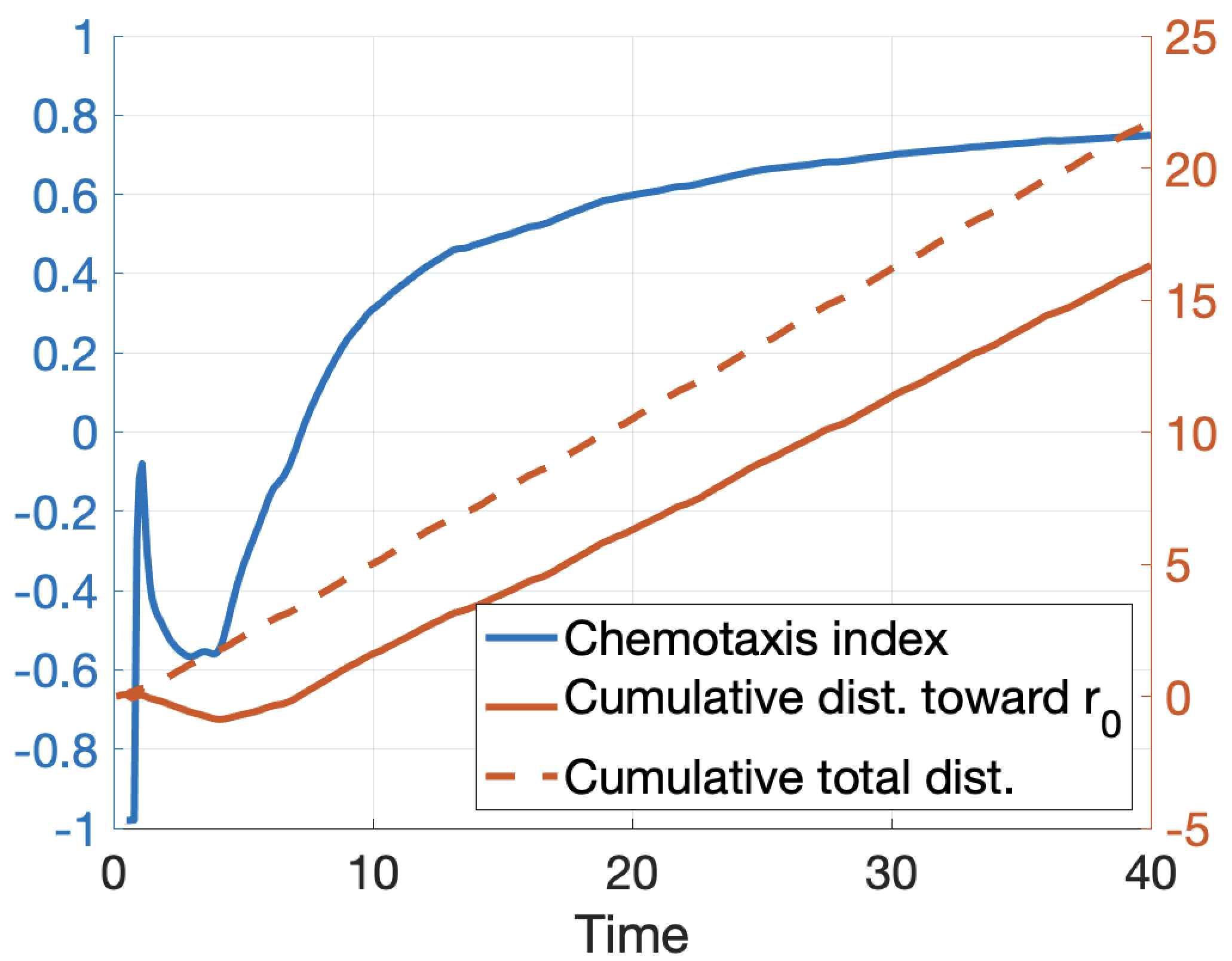}
\includegraphics[width=0.3\linewidth]{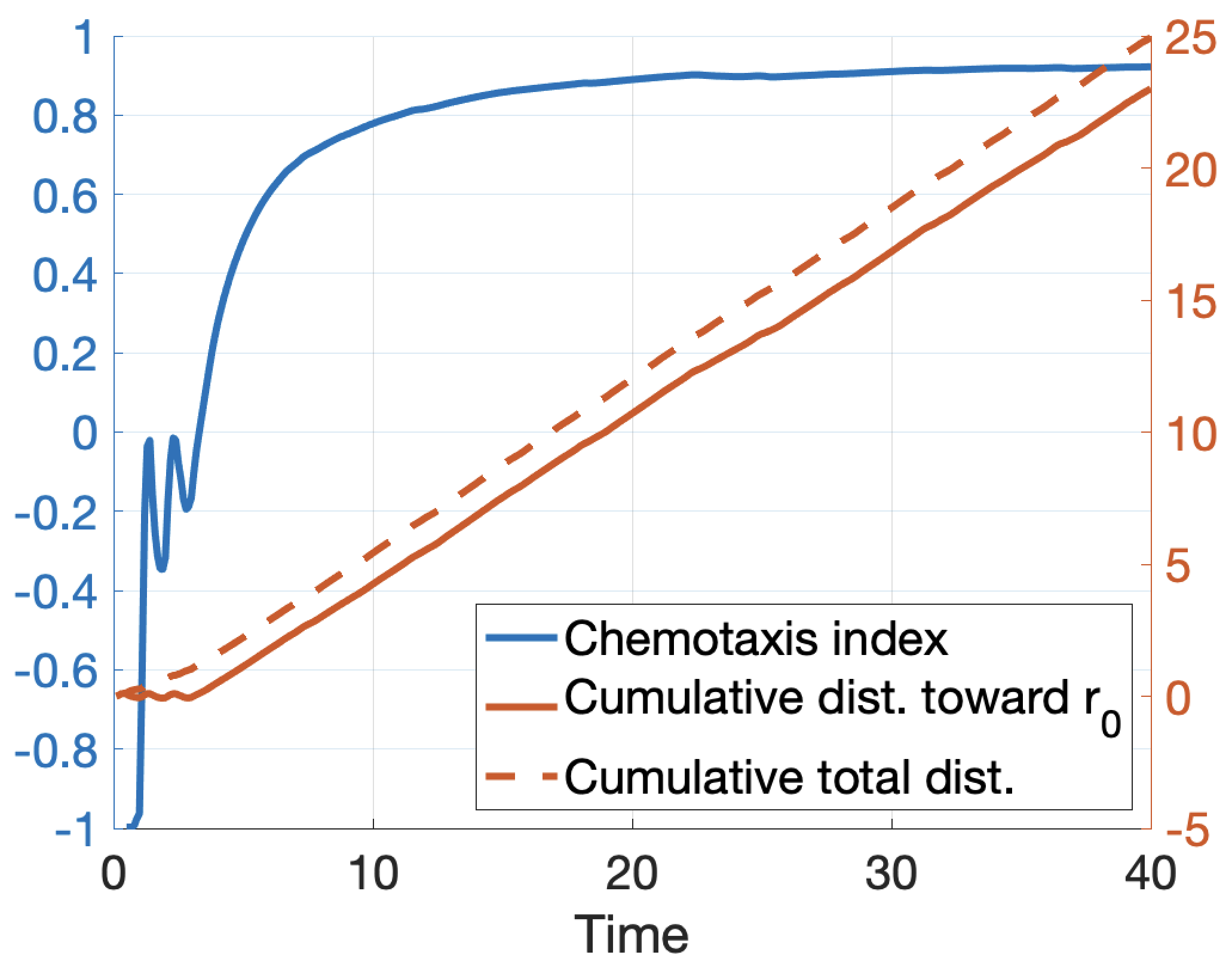} \\
\end{center}
\caption{Chemotaxis indexes for three cell trajectories. Top row: Three cell trajectories for different strengths of chemical protrusion $\alpha = 0.3, 0.4, 0.6$. Bottom row: The corresponding chemotaxis indexes (CI) over time $[0,60]$. As indicated by CI, we note that the greater $\alpha$ becomes, the more directly (and the faster) the cell moves towards the chemoattractant source $\mathbf{r}_0$.   }
\label{fig:Mov03} 
\end{figure}

In Figure \ref{fig:Mov03}, we plot the CI for three simulated cells with various $C_{\text{chem}} = 0.01, 0.03, 0.05$ of the strength of the bias. The top row, from left to right, are the trajectories of three cells up to time $T = 40$, with different values of $C_{\text{chem}} = 0.01, 0.03, 0.05$, respectively. The cells are plotted in the box $[-20,20]\times[-30,10]$, with the chemoattrant source located at $\text{r}_0 = [0,-40]^T$. The bottom row, from left to right, are the CI of the three cells up to time $T = 40$, with different values of $C_{\text{chem}} = 0.01, 0.03, 0.05$, respectively. In each plot of the CI, CI curve is in blue, with the values indicated by the left $y$-axis. The solid and dashed orange curves (values indicated by the right $y$-axis) are the cumulative distance travelled by the cell in the direction toward $\textbf{r}_0$, and the cumulative total distance, respectively.

Our result shows that when the strength of the bias is weaker ($C_{\text{chem}} = 0.01$), the cell wanders more randomly along the trajectory toward the chemoattractant source; while when the strength of the bias becomes stronger ($C_{\text{chem}} = 0.05$), the cell's moving direction is more straightforward. On the other hand, even for the case with a weaker strength of the bias $C_{\text{chem}} = 0.01$, our simulation still shows a high value for CI (CI becomes close to $0.6$ at $t = 40$), indicating efficient cell navigation. For the case with a stronger strength of the bias $C_{\text{chem}} = 0.05$, the CI can reach an even higher value, close to $0.9$.

\section{Discussion and outlook}

Here we have presented a new computational approach to the problem of chemotactically driven motion in Dictyostelium. Our results are consistent with experimental findings and recapitulate results obtained by the level set method \cite{Neilson_PLosBio2011}. While both the level set method and phase field model can couple the Meinhardt dynamics with membrane evolution, thereby successfully reproducing the pseudopod morphology, our method enjoys the fact that the coupling of the phase field with the Meinhardt dynamics is much more straightforward, by simply incorporating $g(\phi)=\frac{\epsilon}{2}|\nabla\phi|^2$ in the Meinhardt equations (\ref{eqn:a}-\ref{eqn:c}). In addition, although it has been claimed that an evolving cell boundary solved by (ALE-SFEM) plus the level set approach has the advantage of efficiency \cite{Neilson_PLosBio2011, Neilson_SISC2011}, such a computational strategy suffers from a very complicated implementation. Indeed, the level-set-modeled cell profile has to “communicate” with the evolving cell boundary at every single time step. More specifically, two set of meshes need to be introduced, the finite element mesh for the cell membrane update, and level set mesh for the update of the level set function. In each time step, one needs to project the finite element mesh points onto level set mesh points by using nearest-neighbor point, in order to update level set function; then use the level set mesh points to form a new finite element mesh, on which the Meinhardt is updated. Since the new finite element mesh may fail to be equidistributed, a step of re-gridding the finite element mesh is often.  In contrast, in our phase field framework, when the cell movement and membrane reaction-diffusion system are combined using the same implicit tracking language, no further “communication” between mechanical and chemical systems is needed, as they are solved on the same uniform mesh (see numerical details in Appendix). 

Finally, in addition to computational efficiency and formulaic complexity, our phase field framework has an obvious advantage when it becomes necessary to couple intra-cellular flow and focal adhesions into the model \cite{Shao_PNAS2012}. In fact, it is definitely worth exploring chemotaxis with a more biophysically complete model, i.e. with all the effects of bulk reaction-diffusion dynamics, cytoplasmic flow, Meinhardt patterns, focal adhesions, and membrane forces. Quite a few other interesting features, apart from bifurcating pseudopods, were observed in experiments more than a decade ago \cite{Alamo_PNAS2007}. For example, when Dictyostelium discoideum cells adhere to the substrate, they exert opposing pole forces that are orders of magnitude higher than required to overcome the resistance from their environment. Also, the strain energy exerted by migrating Dicty on the substrate is (almost) quasi-periodic and can be used to identify different stages of the cell motility cycle. Moreover, the period displays an inversely proportional relation with cell velocity.  In recent work by Copos et al. \cite{Copos_BiophysicalJ2017}, a simple mechanochemical model of 2D (in the vertical plane) cell motility was used to study the periodic changes in cell length and the related spatiotemporal dynamics of traction forces. Our phase field model can provide a platform to carefully study these features in future work.

\section{Acknowledgements}
H. Levine's work is supported by the NSF, grants Nos. PHY-1935762 and PHY-2019745. Y. Zhao’s work is supported by a grant from the Simons Foundation through Grant No. 357963 and NSF grant DMS2142500.

\section*{Appendix}
\label{s:Appendix}
\renewcommand{\thesection}{A}
\setcounter{equation}{0}

In this appendix, we present in detail the numerical algorithm to solve the coupled system (\ref{eqn:a})-(\ref{eqn:c}) and (\ref{eqn:phi}). We take the computational domain $\Omega = [-L_x,L_x)\times[-L_y, L_y)$. Periodic boundary conditions are used for the coupled system. A uniform grid $\Omega_h$ is generated over $\Omega$ by taking $h_x = \frac{2L_x}{N_x}$ and $h_y = \frac{2L_y}{N_y}$. The grid points are given as $(x_i, y_j) = (-L_x+(i-1)h_x, -L_y+(j-1)h_y)$. Given initial data $(\phi^0, a^0, b^0, c^0)$, we aim to find $(\phi^n, a^n, b^n, c^n)$ for $n=1,2,\cdots, N_t$ with $N_t = \frac{T}{\Delta t}$.

For solving the equation of $\phi$ in (\ref{eqn:phi}), we adopt a semi-implicit Fourier spectral method. More specifically, we discretize the equation as
\begin{align*}
\tau\frac{\phi^{n+1}-\phi^{n}}{\Delta t} = &-\kappa\Delta^2\phi^{n+1} - \kappa\nabla^2\frac{G'(\phi^n)}{\epsilon^2} + \kappa \frac{G''(\phi^n)}{\epsilon^2} \left( \nabla^2\phi^n - \frac{G'(\phi^n)}{\epsilon^2}  \right) + \gamma \left( \nabla^2\phi^{n+1} - \frac{G'(\phi^n)}{\epsilon^2}  \right) \nonumber\\
&- M_{\text{area}}\left( \int \frac{\epsilon}{2}|\nabla\phi^{n}|^2 + \frac{1}{\epsilon}G(\phi^n) \text{d}\textbf{r} - P_0 \right)|\nabla\phi^n| + \alpha \tilde{a}|\nabla\phi^n|.
\end{align*}
This discretization can be rewritten as
\[
\left(\frac{\tau}{\Delta t} + \kappa\nabla^4 -\gamma\nabla^2 \right)\phi^{n+1} = \text{RHS}(\phi^n),
\]
which can be efficiently solved by the Fourier spectral method.

Next, we consider the numerical method for the Meinhardt system (\ref{eqn:a})-(\ref{eqn:c}). For the sake of numerical stability, the Meinhardt equations (\ref{eqn:a})-(\ref{eqn:c}) are replaced by:
\begin{align}
&\tau_0 \frac{\partial(\tilde{g}(\phi)a)}{\partial t} + \nabla\cdot(\tilde{g}(\phi)a\mathbf{v}) = D_a\nabla_{\parallel}\cdot(g(\phi)\nabla_{\parallel}a) + D_{\perp}\nabla_{\perp}\cdot(g(\phi)\nabla_{\perp}a) \nonumber\\
&\hspace{2.5in}+ \tilde{g}(\phi)\left( \frac{s(\textbf{r},t)(a^2b^{-1}+b_a)}{(s_c+c)(1+s_aa^2)} - r_a a \right), \label{eqn:a2}\\
&\tau_0 \frac{\partial b}{\partial t}  =  r_b\frac{\int \tilde{g}(\phi)a\text{d}\textbf{r}}{\int \tilde{g}(\phi)\text{d}\textbf{r}} - r_b b , \label{eqn:b2}\\
&\tau_0 \frac{\partial(\tilde{g}(\phi)c)}{\partial t} + \nabla\cdot(\tilde{g}(\phi)c\mathbf{v}) = D_c\nabla_{\parallel}\cdot(g(\phi)\nabla_{\parallel}c) + D_{\perp}\nabla_{\perp}\cdot(g(\phi)\nabla_{\perp}c) + \tilde{g}(\phi)\left( b_c a - r_c c \right), \label{eqn:c2}
\end{align}
in which $\tilde{g}(\phi) = \frac{1}{\epsilon}G(\phi)$.
The replacement of $g$ by $\tilde{g}$ is reasonable due to the fact that in the Ginzburg-Landau functional (\ref{energy:tension}), the term $g(\phi) = \frac{\epsilon}{2}|\nabla\phi|^2$ plays identical role as $\tilde{g}(\phi) = \frac{1}{\epsilon}G(\phi)$ at the system equilibrium \cite{Modica_ARMA1987, LiZhao_SIAP2013}. 

Note that we replace all the terms of $g(\phi)$ by $\tilde{g}(\phi)$ except for those in the parallel and perpendicular diffusion terms. We do this is because by taking $g(\phi) = \frac{\epsilon}{2}|\nabla\phi|^2$ together with the parallel and perpendicular gradient operators (\ref{eqn:par_perp}), the diffusion terms can be significantly simplified. Explicitly, for the parallel and perpendicular diffusion terms in the equation of $a$,
\begin{align*}
\nabla_{\parallel}\cdot(g(\phi)\nabla_{\parallel}a) &=
\begin{bmatrix*}[c]
      n_y^2   & -n_xn_y  \\
      -n_xn_y    & n_x^2 \\
\end{bmatrix*}
\begin{bmatrix*}[c]
      \partial_x  \\
      \partial_y \\
\end{bmatrix*}
\cdot
\left(
g(\phi)
\begin{bmatrix*}[c]
      n_y^2   & -n_xn_y  \\
      -n_xn_y    & n_x^2 \\
\end{bmatrix*}
\begin{bmatrix*}[c]
      \partial_x a  \\
      \partial_y  a\\
\end{bmatrix*}
\right)
\\
&
= \frac{\epsilon}{2}
\begin{bmatrix*}[c]
      n_y^2   & -n_xn_y  \\
      -n_xn_y    & n_x^2 \\
\end{bmatrix*}
\begin{bmatrix*}[c]
      \partial_x  \\
      \partial_y \\
\end{bmatrix*}
\cdot
\begin{bmatrix*}[c]
      (\partial_y\phi)^2\partial_x a - (\partial_x\phi \partial_y\phi) \partial_y a  \\
      - (\partial_x\phi \partial_y\phi) \partial_x a + (\partial_x\phi)^2\partial_y a \\
\end{bmatrix*}
\\
&= \frac{\epsilon}{2} \Bigg[\Big( n_y^2\partial_x - n_xn_y\partial_y \Big)\Big( (\partial_y\phi)^2\partial_x a - (\partial_x\phi \partial_y\phi) \partial_y a \Big) \\
& \qquad\qquad\qquad + \Big( - n_xn_y\partial_x + n_x^2\partial_y  \Big)\Big( - (\partial_x\phi \partial_y\phi) \partial_x a + (\partial_x\phi)^2\partial_y a \Big)\Bigg], 
\end{align*}
and
\begin{align*}
\nabla_{\perp}\cdot(g(\phi)\nabla_{\perp}a) &=
\begin{bmatrix*}[c]
      n_x^2   & n_xn_y  \\
      n_xn_y    & n_y^2 \\
\end{bmatrix*}
\begin{bmatrix*}[c]
      \partial_x  \\
      \partial_y \\
\end{bmatrix*}
\cdot
\left(
g(\phi)
\begin{bmatrix*}[c]
      n_x^2   & n_xn_y  \\
      n_xn_y    & n_y^2 \\
\end{bmatrix*}
\begin{bmatrix*}[c]
      \partial_x a  \\
      \partial_y  a\\
\end{bmatrix*}
\right)
\\
&
= \frac{\epsilon}{2}
\begin{bmatrix*}[c]
      n_x^2   & n_xn_y  \\
      n_xn_y    & n_y^2 \\
\end{bmatrix*}
\begin{bmatrix*}[c]
      \partial_x  \\
      \partial_y \\
\end{bmatrix*}
\cdot
\begin{bmatrix*}[c]
      (\partial_x\phi)^2\partial_x a + (\partial_x\phi \partial_y\phi) \partial_y a  \\
      (\partial_x\phi \partial_y\phi) \partial_x a + (\partial_y\phi)^2\partial_y a \\
\end{bmatrix*}
\\
&= \frac{\epsilon}{2} \Bigg[\Big( n_x^2\partial_x + n_xn_y\partial_y \Big)\Big( (\partial_x\phi)^2\partial_x a + (\partial_x\phi \partial_y\phi) \partial_y a \Big) \\
& \qquad\qquad\qquad + \Big(  n_xn_y\partial_x + n_y^2\partial_y  \Big)\Big(  (\partial_x\phi \partial_y\phi) \partial_x a + (\partial_y\phi)^2\partial_y a \Big)\Bigg]. 
\end{align*}
To numerically discretize the above two terms, firstly we evaluate $(\partial_x\phi,\partial_y\phi)$ using Fourier spectral method, and calculate $(n_x, n_y)$ as 
\[
n_x = \frac{\partial_x\phi}{\sqrt{(\partial_x\phi)^2 + (\partial_y\phi)^2 + \epsilon_0}}, \ n_y = \frac{\partial_y\phi}{\sqrt{(\partial_x\phi)^2 + (\partial_y\phi)^2 + \epsilon_0}},
\]
in which $\epsilon_0$ is a sufficiently small constant (say, $\epsilon_0 = 1e-8$) to avoid dividing zero. Secondly, we evaluate 
\[
\Big(\partial_x((\partial_x\phi)^2), \partial_y((\partial_x\phi)^2)\Big), \Big(\partial_x(\partial_x\phi\partial_y\phi), \partial_y(\partial_x\phi\partial_y\phi)\Big),\Big(\partial_x((\partial_y\phi)^2), \partial_y((\partial_y\phi)^2)\Big)
\] using Fourier spectral method. Thirdly, the first and second derivatives of $a$ are evaluated by central difference:
\begin{align*}
&\partial_x a \approx \frac{a_{i+1,j}-a_{i-1,j}}{2h_x}, \ \partial_y a \approx \frac{a_{i,j+1}-a_{i,j-1}}{2h_y}, \ \partial_{xx}a \approx \frac{a_{i+1,j}-2a_{ij}+a_{i-1,j}}{h_x^2},\\
&\partial_{xy}a \approx \frac{a_{i+1,j+1}-a_{i-1,j+1}-a_{i+1,j-1}+a_{i-1,j-1}}{4h_xh_y}, \ \partial_{yy}a \approx \frac{a_{i,j+1}-2a_{ij}+a_{i,j-1}}{h_y^2}.
\end{align*}
Inserting all evaluations above back, we obtain the numerical approximation of $\nabla_{\parallel}\cdot(g(\phi)\nabla_{\parallel}a)$ and  $\nabla_{\perp}\cdot(g(\phi)\nabla_{\perp}a)$.

The advection term $\nabla\cdot(\tilde{g}(\phi)a\mathbf{v})$ in the equation of $a$ is approximated by a central difference scheme:
\begin{align*}
\nabla\cdot(\tilde{g}(\phi)a\mathbf{v})& \approx \frac{ \tilde{g}(\phi_{i+\frac{1}{2},j}) a_{i+\frac{1}{2},j}v^x_{i+\frac{1}{2},j} - \tilde{g}(\phi_{i-\frac{1}{2},j}) a_{i-\frac{1}{2},j}v^x_{i-\frac{1}{2},j} }{h_x}  \\
&\qquad+ \frac{ \tilde{g}(\phi_{i,j+\frac{1}{2}}) a_{i,j+\frac{1}{2}}v^y_{i,j+\frac{1}{2}} - \tilde{g}(\phi_{i,j-\frac{1}{2}}) a_{i,j-\frac{1}{2}}v^y_{i,j-\frac{1}{2}} }{h_y}
\end{align*}
in which  $\mathbf{v} = [v^x,v^y]^T = -\partial_t\phi\frac{\nabla\phi}{|\nabla\phi|^2}$ and is calculated by taking $\partial_t\phi \approx \frac{\phi^{n+1} - \phi^n}{\Delta t}$, and $\nabla\phi$ evaluated by Fourier spectral approximation.

The time derivative $\frac{\partial(\tilde{g}(\phi)a)}{\partial t}$ is approximated by  forward Euler scheme,
\[
\frac{\partial(\tilde{g}(\phi)a)}{\partial t} \approx \frac{ \tilde{g}(\phi^{n+1})a^{n+1} - \tilde{g}(\phi^{n})a^{n} }{\Delta t} .
\]
Finally, with all terms discretized in the equation of $a$, we get an update on $a$: $a^n \rightarrow a^{n+1}$.  The equation (\ref{eqn:c2}) can be solved numerically in a similar manner. The equation (\ref{eqn:b2}) is an ODE, so we can adopt an efficient fourth-order Runge-Kutta method (RK4) to solve it.

Since the phase field cell $\phi$ moves around in the computational domain $\Omega$ and may near the edge, we do not solve the phase field equation and the Meinhardt equations in the entire domain. We only solve these equations in a smaller box of size $1.75L_x\times 1.75L_y$ near the cell. This box is re-centered if the cell is close to one of its four boundaries: if $\phi\ge 0.5$ within $\frac{N_x}{16}$ (or $\frac{N_y}{16}$) pixels of the boundary, the box is shifted $\frac{N_x}{4}$ (or $\frac{N_y}{4}$) pixels away from the boundary. We treat the small box as having periodic boundary conditions, which is appropriate as we keep the cell from too closely approaching the edge of $\Omega$.



\end{document}